\documentclass[aps,prd,nofootinbib,longbibliography, 10pt]{revtex4-1}
\usepackage{amsmath}
\usepackage[T1]{fontenc}
\usepackage{amsfonts}
\usepackage{amssymb}
\usepackage{dsfont}
\usepackage{mathrsfs}
\usepackage[mathscr]{euscript}
\usepackage{yhmath}
\usepackage{epstopdf}
\usepackage{multirow}
\usepackage{caption}
\usepackage{tensor}
\usepackage{enumerate}
\usepackage{enumitem}
\usepackage{subcaption}
\usepackage{color}
\usepackage{graphicx}
\usepackage{dcolumn}
\usepackage{bm}
\usepackage{natbib}
\usepackage{float}
\usepackage{hyperref}
\hypersetup{
    colorlinks=true, 
    linktoc=page,    
    linkcolor=blue,  
    urlcolor=magenta
}

\allowdisplaybreaks

\newcommand{\va}{\scriptscriptstyle}
\newcommand{\be}{\nopagebreak[3]\begin{equation}}
\newcommand{\ee}{\end{equation}}

\newcommand{\ba}{\nopagebreak[3]\begin{eqnarray}}
\newcommand{\ea}{\end{eqnarray}}

\newcommand{\la}{\label}
\newcommand{\n}{\nonumber}
\newcommand{\su}{\mathfrak{su}}
\newcommand{\SU}{\mathrm{SU}}
\newcommand{\U}{\mathrm{U}}

\begin{document}

\title{Asymptotically de Sitter universe inside a Schwarzschild black hole}

\author{Emanuele Alesci$^{1,2}$}
\email{emanuele.alesci@gmail.com }
\author{Sina Bahrami$^2$}
\email{sina.bahrami.igc@gmail.com}
\author{Daniele Pranzetti$^3$}
\email{dpranzetti@perimeterinstitute.ca}
\affiliation{$^1$Zhejiang University of Technology (ZJUT)
18 Chaowang Road, Hangzhou, Zhejiang, 310014, China}
\affiliation{$^2$Institute for Gravitation and the Cosmos, Pennsylvania State University, University Park, Pennsylvania 16802, USA}
\affiliation{$^3$Perimeter Institute for Theoretical Physics,  Waterloo, Ontario N2L 2Y5, Canada}

\date{\today}

\begin{abstract}
Extending our previous analysis, we study the interior of a Schwarzschild black hole derived from a partial gauge fixing of the full loop quantum gravity Hilbert space, this time including the inverse volume and coherent state subleading corrections.  
Our derived effective Hamiltonian differs crucially from the ones introduced in the minisuperspace
models. This distinction is reflected in the class of  homogeneous bouncing geometries that replace the classical singularity and are labeled by a set of quantum parameters associated with the structure of  coherent states used to derive the effective Hamiltonian.
By fixing these quantum parameters through geometrical considerations,
the post-bounce interior geometry reveals a high sensitivity to the value of the Barbero--Immirzi parameter $\gamma$. Surprisingly, we find that  
$\gamma\approx 0.274$ results in an asymptotically de Sitter geometry in the interior region,
where now a cosmological constant is generated purely from quantum gravitational effects. 
The striking fact is the {\it exact} coincidence of this value for $\gamma$ with the one derived from the $\SU(2)$ black hole entropy calculations in loop quantum gravity. The emergence of this value in two entirely unrelated theoretical frameworks and computational setups is strongly suggestive of deep ties between the area gap in loop quantum gravity, black hole physics, and the observable universe. In connection to the latter, we point out an intriguing relation between the measured value of the cosmological constant and the observed mass in the universe from a proposal for a spin quantum number renormalization effect associated to the microscopic dynamics.
\end{abstract}

\maketitle

\tableofcontents

\section{Introduction}

It is often stated that black holes are the hydrogen atoms of quantum gravity; just as the classical instability of hydrogen atom signified the need for a quantum description of the subatomic interactions, the formation of spacetime singularities as predicted classically in \cite{Penrose:1964wq, Hawking:1969sw} as the final stage of gravitational collapse reflects the need for a UV completion of general relativity (GR), where quantum properties of geometry are widely expected to circumvent these pathologies.  
Similarly, as spectroscopy of the hydrogen atom provided one of the first experimental tests of the quantum theory, the recent advances in detection of gravitational waves \cite{Abbott:2016blz} and interferometric imaging \cite{Akiyama:2019cqa} are expected to provide a possible experimental window into the nature of spacetime at the Planck scale by revealing quantum properties of the near horizon geometry (see, e.g.,  \cite{Giddings:2016btb, Giddings:2019jwy,  Abedi:2020ujo}). In fact, this analogy has  motivated an ``atomistic'' approach to black hole evaporation
\cite{Bekenstein:1995ju, Krasnov:1997yt, Barrau:2011md, Pranzetti:2012pd, Pranzetti:2012dd}, with the Hawking radiation spectrum consisting of discrete emission lines between different horizon area/energy levels. It is therefore no exaggeration to say that just as the hydrogen atom represented a system simple enough but at the same time of great physical relevance for testing the predictions of quantum mechanics, the Schwarzschild black hole provides the perfect arena to apply the formalism of a given quantum gravity theory and explore its theoretical and possibly observational consequences. 

loop quantum gravity (LQG) provides a nonperturbative, background independent quantization of GR as formulated in the Ashtekar connection variables
\cite{Thiemann:2007zz}. As of now, it represents one of the most advanced programs for UV completion of GR. In particular, its canonical formulation is perfectly tailored to the study of singularity resolution both in cosmology and black hole physics. However, while the canonical LQG quantum dynamics can be  formulated through the rigorous definition of a Hamiltonian constraint operator \cite{Thiemann:1996aw} (though not in an ambiguity-free scheme), the study of its solutions is a formidable task \footnote{In addition to the known ambiguities in the scalar Hamiltonian constraint, there is the problem of having no finite generator for the spatial diffeomorphism constraint in LQG. Presently, the concepts of covariant dynamics and spacetime gauge transformations are not well understood in this theory. See \cite{Laddha:2011mk, Tomlin:2012qz,PhysRevD.97.106007} for recent progress in this direction. Nevertheless, we should emphasize that the covariance issue self-resolves if the spacetime under study is homogeneous, which is the case for the Schwarzschild interior.}. This technical complexity is not surprising though, given that even classically no general solution of the Einstein equations is known. Nonetheless, concrete calculations can be carried out when symmetry assumptions are made to simplify the equations.

Implementing a symmetry reduction classically is clear and straightforward. 
Relevant to our current study is the spherical symmetry reduction of the phase space which yields a minisuperspace where only a finite number of degrees of freedom remain and the constraints simplify greatly. After specifying a 3+1 decomposition for the spacetime, one can write down the evolution equations for the minisupersapce set of conjugate variables.  One then solves the simplified dynamics for a given choice of initial data that serves to fix the independent constants of integration often by invoking physical considerations for the system under study. We will review the Hamiltonian approach to the Schwarzschild interior in Section \ref{sec:CPP}. 

However, this process becomes much more subtle when passing to the quantum theory. In fact, there are several strategies to perform the symmetry reduction and, in general, these lead to dynamics capturing a different number/typologies for the degrees of freedom. The main choice to be made is whether to impose symmetry at the classical or at the quantum level. The first path leads to the minisuperspace  models, where one quantizes the reduced phase space within the event horizon using the techniques developed in loop quantum cosmology (LQC)  \cite{Ashtekar:2011ni} as this region is a Kantowski-Sachs cosmological spacetime.  This line of investigation was started in \cite{Modesto:2005zm, Ashtekar:2005qt} and more recently continued in \cite{Gambini:2013hna}. While these studies provided the first evidence for the Schwarzschild singularity resolution at the quantum level, they fell short of depicting a robust picture for the post-singularity geometry. Moreover, due to the inhomogeneity of the Schwarzschild spacetime and the use of point holonomies erasing the graph structure on the 2-spheres foliating the homogeneous leaves in the interior, issues related to gluing of the interior with the exterior geometry and any potential link to the full quantum theory remained open. This has motivated an {\it effective geometry} approach where one introduces a modified Hamiltonian constraint on the classical phase space encoding some quantum geometry effects expected from an LQG treatment.  By solving the dynamical flow generated by this effective Hamiltonian, it was shown in \cite{Ashtekar:2018cay} that an antitrapped region emerges to the future of the 3D space-like  transition surface replacing the classical singularity. A proposal for extending this effective analysis to the exterior region was also presented \footnote{See \cite{Ashtekar:2020ckv}  for a discussion on some of the deviations from the classical asymptotic properties of the  spacetime metric.}. An alternative approach to effective dynamics within polymer models has been followed in
\cite{Bojowald:2015zha, BenAchour:2017jof, BenAchour:2018khr, Bojowald:2018xxu, Aruga:2019dwq}. In this case, the quantum corrections to the Hamiltonian constraint are dictated by the requirement to preserve (a modified version of) general covariance for the theory, namely to maintain even at the effective level a deformed but closed off-shell constraint algebra. In this class of polymer models one finds a signature change for the effective metric from Lorentzian to Euclidean inside the trapped region. The relation to and derivation from the full LQG theory remain the key open issues in all of these  minisuperspace models. 

It should be noted that the status of this effective approach to the black hole interior dynamics is even murkier than in the cosmological case. In the latter, the effective expression for the Hamiltonian constraint  initially introduced to study the resolution of the big bang singularity was eventually derived \cite{Taveras:2008ke} from the expectation value of the polymerized quantum Hamiltonian on sharply peaked states. In the black hole case, however, this derivation was not achieved.

Significant progress has recently been made in an alternative approach to a simplified dynamics. This entails the technically more involved choice of starting with the quantization of the full phase space of the theory and only later performing the symmetry reduction at the quantum level
\footnote{ 
A different framework for the definition of  continuum spherically symmetric quantum geometries within the full theory is provided by the Group Field Theory reformulation of LQG in a second quantization language  \cite{Oriti:2015qva, Oriti:2018qty}.  In this case, spherical symmetry is encoded in the homogeneity properties of  the  condensate wave-functions used in the construction of a generalized class of coherent states peaked on some global geometrical observables.}. This `quantum reduced loop gravity' (QRLG) framework was originally developed  for the cosmological applications \cite{Alesci:2013xd, Alesci:2014rra, Alesci:2014uha, Alesci:2017kzc, Alesci:2016xqa}, and  later refined and extended to  study spherically symmetric black holes  \cite{Alesci:2018ewg, Alesci:2018loi}. This approach involves two main steps; the first one consists of the imposition of a partial gauge fixing of the full kinematical Hilbert space by reducing both the spin network states and the holonomy-flux algebra operators entering the construction of the constraints---very recently, it has been shown that indeed this construction is equivalent to the action of the full theory operators on the reduced kinematical Hilbert space in the limit of  large spin quantum numbers \cite{Makinen:2020rda}, further consolidating the  technical foundations of the framework---. This partial gauge fixing is a necessary step for introducing coherent states peaked around a spherically symmetric classical geometry to be used to compute the expectation value of the Hamiltonian constraint operator. Completion of this second step yields the sought after effective Hamiltonian defining the  quantum corrected dynamics for a  spherically symmetric  geometry. This program is reviewed in Section \ref{sec:eff-H}, where we also supplement the previous construction in \cite{Alesci:2018loi} with the inclusion of 
the first sub-leading terms from the  inverse volume corrections, 
  as well as coherent state corrections in the spread parameters' expansion.
As our analysis does not rely on polymer quantization techniques, we do not have to restrict to a homogenous foliation from the beginning. In fact, the general form of the effective Hamiltonian that we derive is for a horizon penetrating foliation and it can be used to evolve an initial data on both the interior and the exterior regions at the same time (barring the complexities caused by lacking the effective analog of the spatial diffeomorphism constraint). However, solving the lengthy and non-local scalar constraint equation in this general case is a daunting task that requires advanced numerical techniques. While this is currently under investigation, here we are mainly concerned with the asymptotic geometry of the post-bounce extension of the spacetime and we can thus restrict to the interior homogenous foliation to simplify the expression of the effective Hamiltonian. This is  done in Section \ref{sec:interior}, where we show how the quantum corrected dynamics  of the interior black hole region investigated in \cite{Alesci:2019pbs} can be derived. 
It is important to keep in mind that the restriction to a homogeneous foliation is just a  simplification  used {\it a posteriori}, once the full LQG machinery has been deployed. This has crucial implications, as we explain in what follows.

As in the previous minisuperspace models, the resulting expression for the effective Hamiltonian depends on the quantum parameters codifying the discrete regularization structure underlying the construction of the kinematical Hilbert space. However, having included the geometrical data associated to the  full graph structure from the beginning, we are now able to fix the dependence of these quantum parameters on the phase space variables through clear geometrical considerations, thus considerably reducing the ambiguities affecting the final predictions of the theory. The second major difference with previous analyses has also its origin in the inclusion of the extended geometrical data ($\SU(2)$ link holonomies instead of point holonomies) on the 2-spheres foliating the leaves. More precisely, integration over the 2-sphere angular coordinates, which amounts to averaging out fluctuations around spherical symmetry in the effective theory instead of freezing them out from the get-go as in LQC-like treatments, yields the appearance of the Struve function (of zeroth order) in the effective Hamiltonian. As shown in  \cite{Alesci:2019pbs}, it is the different (non-periodic and decaying) behavior of the Struve function from the sine function used in the polymer models which prevents the formation of a white hole horizon in the antitrapped region (of the effective spacetime metric) to the future of the moment of bounce. Going back to the hydrogen atom analogy for a moment, it is suggestive to think of the zeroth order Struve function as the imaginary counterpart of the  Bessel function of the first kind \footnote{They represent respectively the imaginary and the real parts of the integral $\int_0^\pi e^{i x\sin{\theta}} d\theta$ that appears from some holonomic components of the Hamiltonian constraint.}, as this is used to describe the bound states of an electron in a hydrogen-like atom.

Therefore, while the analysis in  \cite{Alesci:2019pbs} revealed the importance of equally treating  holonomies in all directions in order to capture the correct essential features of the full theory dynamics, it also exhibited some properties of the effective metric solution found in \cite{Ashtekar:2018cay}. More precisely, it confirmed that at the moment of bounce all curvature invariants have a mass-independent upper bound and no large quantum effects are present near the classical event horizon (as long as the black hole's mass far exceeds the Planck mass). 
Moreover,  the transition surface where the bounce occurs is located in proximity of the spacetime region where the curvature (and not the radius of 2-spheres) becomes Planckian (this is an important point that we will elaborate more on in Section \ref{sec:CC}). 

As stressed above, our approach differs substantially  from the polymer models in the post-bounce behavior of the effective metric.  As shown in  \cite{Alesci:2019pbs}, while the fine features depend on the choice of some quantum parameters entering the construction of the coherent states, the class of solutions obtained from the effective evolution equations matched the geometry of a homogeneous expanding universe, with no finite distance boundary in the antitrapped region. However, the rates of expansion for the two metric functions describing the effective spatial geometry depend on the numerical value of the parameter
$\eta:= {\alpha}/{\beta}$,
where $\alpha$ and $ \beta $ are two constants with dimension of length that depend on two (averaged) quantum spin numbers and the Barbero--Immirzi parameter $\gamma$.
In Section \ref{sec:eff-H} we will improve our construction in  \cite{Alesci:2019pbs} by implementing an extra geometrical condition on the coherent states descending from the covariant formulation of the full theory. This will allow us to reduce the ambiguities in the solution space by limiting the dependence of $\eta$  on $\gamma$ only. 
In this way, it is the Barbero--Immirzi parameter that uniquely determines the properties of the leading term (in a near infinity expansion) of the post-bounce asymptotic metric, as explicitly derived in Section \ref{sec:eff-eq}.

At this point we are ready to ask the main question addressed in this paper: Is there a value of the Barbero--Immirzi parameter for which the post-bounce geometry becomes asymptotically de Sitter, as defined in \cite{Ashtekar:2014zfa}? 

The answer to this question is carefully worked out in Section \ref{sec:dS}. We first perform a series expansion of the phase space variables near the post-bounce asymptotic infinity, where we match the leading terms of the metric functions with those of the de Sitter metric expressed in a coordinate system adapted to the Schwarzschild interior homogeneous foliation (this is briefly reviewed  in Appendix \ref{App:deSitter}). By expanding the evolution equations to the relevant order of approximation, we derive a set of algebraic equations that are solved by fixing all the free parameters of the theory, up to a free remaining quantum spin number. We will show how including the spread parameter corrections, representing the coherent state first sub-leading terms, play a crucial role in guaranteeing that all the correct requirements of an asymptotically de Sitter geometry are satisfied. Then, going back to the main question, our analysis shows that the demand for the formation of an asymptotically de Sitter universe inside a Schwarzschild black hole selects the following numerical value for the Barbero--Immirzi parameter 
\be\la{Imm}
\gamma\approx 0.274\,. 
\ee
It is an extraordinary fact that this is {\it exactly the same} value required by the LQG $\SU(2)$ black hole entropy calculation  \cite{Engle:2010kt, Agullo:2010zz} in order to obtain the famous factor of 1/4 in the Bekenstein--Hawking entropy-area law \cite{Bekenstein:1973ur, Hawking:1974sw}. As long as the near horizon geometry is sufficiently classical,  this result is robust for the chosen quantum parameters as we demonstrate by a complementary analysis reported in Appendix \ref{sec:AppB}.

To summarize, the dynamical picture is as follows. The metric functions follow the classical dynamical trajectory until the spacetime curvature becomes Planckian. At this point quantum gravity effects become dominant and they manifest themselves in the form of a negative energy density and pressure, which violate the dominant energy condition and catapult the effective dynamical trajectory to a different region of the phase space. At the bounce, all curvature invariants are bounded from above and the singularity is resolved, as further corroborated by the vanishing of both the ingoing and outgoing expansions of the two future directed null normals to the 2-spheres foliating the leaves. After the bounce, a new spacetime antitrapped region opens up, whose geometrical structure is intimately connected with the presence of an area gap in the LQG description of quantum geometry. 
More precisely, while the origin of the bounce can be traced back to a non-zero value for the Barbero--Immirzi parameter in the quantum theory, it is its exact numerical value that determines the asymptotic properties of the  post-bounce effective geometry. The special value \eqref{Imm} plays two different  physical roles in two separate regions. On the one hand, it guarantees the consistency of the quantum description of macroscopic horizons with the semi-classical results of QFT on a fixed curved background. On the other hand, it precisely fine tunes the effective trajectory to evolve into an asymptotically de Sitter universe, revealing the purely quantum gravitational origin of the corresponding positive cosmological constant.

The question of whether this specific value for the area gap and the ensuing implications within the  framework just described can provide an alternative viable ``black hole cosmology'' scenario with possible experimental tests clearly depends on the value of the  cosmological constant in the quantum de Sitter universe it gives birth to. While we intend to address this intriguing scenario in a separate work, we point out in Section \ref{sec:CC} how arguments coming from the consistency of the semi-classical limits are not very helpful in narrowing down an order-of-magnitude estimate for the emerging cosmological constant. Rather, a better control over the microscopic dynamics seems necessary. In fact, if we assume some spin renormalization properties for the quantum geometry evolution, an intriguing estimate for the value of the cosmological constant can be made.


We conclude with a final discussion in Section \ref{sec:disc}.

\section{Review of the Hamiltonian formalism for the Schwarzschild interior}\la{sec:CPP}

Our aim in this paper is to solve the effective 
Hamilton's equations for the interior of the Schwarzschild black hole. Before delving into quantization, the reader likely benefits from a concise discussion of the classical framework. Some of the issues that are discussed below, such as gauge freedom and symmetries,
 will be relevant for the subsequent discussions. 

The Schwarzschild metric has four Killing vector fields; one translational vector field that becomes time-like near $\mathcal{I}$, and three others associated with spherical symmetry. The translational Killing vector field becomes space-like inside the black hole horizon. Since its integral curves are isometric to $\mathbb{R}$, the interior geometry is naturally equipped with a spatially homogeneous foliation. 
The metric in this region can be written as 
\be  \la{4m}
g_{ab}dx^a dx^b = - N(\tau)^2 d \tau^2 + \Lambda(\tau)^2 dx^2 + R(\tau)^2 d \Omega^2,
\ee
where $N$ is the lapse function that determines the foliation and $d \Omega^2 = d \theta^2 + \sin^2{\theta} \ d \phi^2$ is the unit 2-sphere metric. Note that we have omitted the shift vector $\vec{N} = N^x \partial_x$ since it can always be eliminated by
a coordinate transformation of the form 
\be 
\tau \mapsto \tau, \ \ \ x \mapsto x - \int d \tau N^x(\tau).
\ee
Given the symmetries, it is easy to check that the spatial diffeomorphism constraint is identically zero, a fact that bodes well for the self-consistency of the effective covariant dynamics as it relates to the interior geometry.

The Einstein--Hilbert action adapted to metric \eqref{4m} reduces to
\be \la{actclass}
S[N, R, \Lambda]=\frac{1}{G} \int d x \int d \tau \bigg[ \frac{N \Lambda}{2} - \frac{1}{N}\Big(\frac{\Lambda}{2} \dot{R}^2 + R \dot{R} \dot{\Lambda} \Big) \bigg] := \int d\tau \ \mathcal{L}_{\rm c}[N, R, \Lambda, \dot{R}, \dot{\Lambda}],
\ee
where $\mathcal{L}_{\rm c}$ is the classical Lagrangian, dot denotes differentiation with respect to $\tau$, and we have performed the angular integrals and ignored the boundary terms \footnote{Classical quantities are from now on denoted with subscript ${\rm c}$ and we work in $c = \hbar = 1$ units.}. In order to have a well-defined action principle, we regulate the $x$-integral by requiring $x \in [-\mathfrak{L}_0, \mathfrak{L}_0]$.  We define the properly rescaled momenta conjugate to $R$ and $\Lambda$  by 
\be \la{classmom} 
P_R := \frac{1}{2 \mathfrak{L}_0} \frac{\delta S}{\delta \dot{R}} = - \frac{\Lambda \dot{R}+\dot{\Lambda} R}{G N}, \hspace{2 cm} P_{\Lambda} := \frac{1}{2 \mathfrak{L}_0} \frac{\delta S}{\delta \dot{\Lambda}} = - \frac{ R \dot{R}}{G N}.
\ee
The classical scalar Hamiltonian $\mathcal{H}_{\rm c}$ is then obtained by performing a Legendre transform on the classical Lagrangian given in Eq. \eqref{actclass} using the momenta defined above. A straightforward calculation gives 
\be \label{classham}
\frac{\mathcal{H}_{\rm c}}{2 \mathfrak{L}_0} = - \frac{G P_R P_\Lambda}{R} + \frac{G \Lambda P_\Lambda ^2}{2 R^2} - \frac{\Lambda}{2 G}\,.
\ee

It should be clear to the reader by now that the dynamical phase space is parameterized by $R$, $\Lambda$, $P_R$, and $P_\Lambda$. We require them to satisfy the following rescaled Poisson brackets relations:
\be
 \{R, P_R\}=\{\Lambda, P_\Lambda\}=\frac{1}{2 \mathfrak{L}_0}\,.
\ee
The $1/2\mathfrak{L}_0$ factor has been introduced to avoid the divergence in the symplectic structure that arises from the $x$-integral. In contrast to what is usually done in the LQC minisuperspace quantization approach, we do not absorb the $\mathfrak{L}_0$ factors in the phase space variables. The resulting Hamilton's evolution equations that appear below are independent of any fiducial cutoffs, rendering all physical quantities derived in the rest of our analysis  invariant under a rescaling by $\mathfrak{L}_0$.

If the dynamical equations are integrable, conserved quantities are expected to exist. In a constrained Hamiltonian system like general relativity, a conserved quantity (or a Dirac observable) $f$ satisfies the following equation: 
\be \la{diracobseq}
\{f, \mathcal{H}_{\rm c} \} \approx 0.
\ee 
Here $\approx$ denotes evaluation on the constraint surface. In general, solving Eq. \eqref{diracobseq} is challenging since it requires disentangling a non-linear partial differential equation. Nevertheless, the following two independent solutions can be found for the classical Hamiltonian \eqref{classham}:
\be \la{classO}
f_1 = \frac{P_\Lambda ^2}{R}+\frac{R}{G^2}, \hspace{2cm} f_2=R P_R - \Lambda P_\Lambda.
\ee
It turns out that there are no additional Dirac observables except those that can be trivially related to the above expressions by factors of $\mathcal{H}_{\rm c}$. Moreover, it can be shown  that both $f_1$ and $f_2$ are  proportional to the black hole ADM mass $m$ when evaluated along the dynamical trajectories (see the classical solutions given in Eq. \eqref{solc} below). For the 
choice of lapse function given in Eq. \eqref{Nc} below, they become $2 m/G$ and $m$ respectively. A quick calculation shows that $\{f_1,f_2\}=f_1/2 \mathfrak{L}_0$, from which it follows that the ratio $f_2/f_1$ is conjugate to $f_1$. These two quantities are now independent Dirac observables. Unlike $f_2$, $f_2/f_1$ does not have a straightforward physical interpretation. We refer the interested reader to  \cite{Kuchar:1994zk} for further discussion.

\subsection{Hamilton's equations}

In order to construct the interior geometry, we first have to select an initial hypersurface $\Sigma$ in  vicinity of the black hole's event horizon. This is where an appropriate initial data set has to be specified. In particular, the initial data set would need to solve the scalar constraint equation that is given by the vanishing of $\mathcal{H}_{\rm c}$ on $\Sigma$. By virtue of the dynamical equations, it is then straightforward to show that $\mathcal{H}_{\rm c}$ vanishes everywhere along the dynamical trajectories.

The Hamilton's evolution equations are given by 
\begin{subequations} \la{hamileq}
\ba
\dot{R}_{\rm c} &=& \{R_{\rm c}, H_{\rm c}[ N_{\rm c}]\}\,, \\ 
 \dot{\Lambda}_{\rm c}&=& \{\Lambda_{\rm c}, H_{\rm c}[ N_{\rm c}] \}\,, \\
\dot{P}_{R_{\rm c}}&=&  \{P_{R_{\rm c}}, H_{\rm c}[ N_{\rm c}]\}\,,\\
 \dot{P}_{\Lambda_{\rm c}}& =&  \{P_{\Lambda_{\rm c}}, H_{\rm c}[ N_{\rm c}] \}\,,
\ea
 \end{subequations}
where $H_{\rm c} = \int d\tau \  N_{\rm c} \mathcal{H}_{\rm c}$ is the smearing of $\mathcal{H}_{\rm c}$ with a lapse function $ N_{\rm c}$. A choice of $ N_{\rm c}$ corresponds to a rescaling of the proper time by $1/ N_{\rm c}$. It signifies the only gauge freedom in this dynamical system. Note that due to the vanishing of $\vec{N}$ in this symmetry reduced sector of geometry, the constraint algebra is one-dimensional and hence trivial.  

For a black hole of mass $m$, choosing 
\be \label{Nc}
 N_{\rm c} = - \frac{R^2}{2 G^2 m P_{\Lambda}}\,
\ee
yields the following equations:
\begin{subequations}
\ba 
\dot{R}_{\rm c}&=& \frac{ R_{\rm c}}{2 G m}\,,\\
\dot{\Lambda}_{\rm c}&=& -\frac{\Lambda_{\rm c}(R^2 _{\rm c} + G^2 P^2_{\Lambda_{\rm c}})}{4 G^3 m P^2_{\Lambda_{\rm c}}}\,,\\
\dot{P}_{R_{\rm c}}&=&\frac{G^2 P_{R_{\rm c}} P_{\Lambda_{\rm c}}+R_{\rm c} \Lambda_{\rm c}}{2 G^3 m P_{\Lambda_{\rm c}}}\,,
\\
 \dot{P}_{\Lambda_{\rm c}}&=&\frac{-G^2 P^2_{\Lambda_{\rm c}} + R^2 _{\rm c}}{4 G^3 m P_{\Lambda_{\rm c}}}\,.
\ea
\end{subequations}
These can be explicitly integrated to give the following phase space trajectories:
\begin{subequations}\label{solc}
 \ba 
 R_c(\tau)& =& 2 G m \ e^{\tau/2 G m}, \\
\Lambda_c(\tau) &=&\pm \sqrt{e^{- \tau/2 G m} -1}, \\
P_{R_c}(\tau) &=& \frac{1}{2 G} \Big[2- e^{- \tau/2 G m}\Big], \\
 P_{\Lambda_c}(\tau)& =& \mp 2 m \ e^{\tau/ 4 G m} \sqrt{1- e^{\tau/2 G m}}\,.
\ea
\end{subequations}
The proper time defined by the lapse function \eqref{Nc} covers the entire black hole  interior region for the range $-\infty <\tau <0$, where 
$\tau=0$ corresponds to the black hole's event horizon and $\tau = - \infty$ corresponds to the classical singularity. This choice of lapse function  is simply motivated by consistency with our previous work \cite{Alesci:2019pbs}, as it is 
the $\hbar \rightarrow 0$ limit of a different lapse function that drastically simplified the analysis of the quantum-corrected Hamilton's equations for a specific class of coherent states analyzed there. 

Finally, it will be useful to have a quantity that can help differentiate between the classical and quantum regimes.  The  Kretschmann scalar, which is a gauge invariant measure of the spacetime curvature, can be relied on for this task. For the  Schwarzschild metric expressed in the $\{\tau,x,\theta,\phi\}$ coordinates, it becomes
\be\la{K}
\mathcal{K}_c:= R_{abcd} R^{abcd}=\frac{3 \ e^{- \frac{3 \tau}{ G m}}}{4 (G m)^4}\,.
\ee  
The transition to the high curvature regime is signaled by crossing the value of time $\tau \sim \tau_\star$ when the Kretschmann scalar becomes Planckian, namely when $\mathcal{K}_c \sim 1/\ell_{\va P}^4$. $\tau_\star$ is easily found to be
\be\la{tstar}
\tau_\star \sim \frac{G m}{3} \log{\left[\frac{3 \ell_{\va P} ^4 }{4 G^4 m^4}\right]}\,,
\ee
which corresponds to 
\be
R_c(\tau_\star)\sim (Gm)^{\frac{1}{3}} (\ell_{\va P})^{\frac{2}{3}}\,.
\ee

\section{Effective Hamiltonian from Quantum Reduced Loop Gravity}\la{sec:eff-H}

The first  derivation of an effective Hamiltonian constraint for a spherically symmetric geometry starting from the full LQG framework  was performed  in \cite{Alesci:2018loi}. There we extended the QRLG approach that was previously developed for the cosmological case by first  implementing a partial gauge fixing of the LQG kinematical Hilbert space compatible with the construction of coherent states peaked around spherically symmetric geometrical data. By defining the partially gauge fixed holonomy-flux algebra operators, we constructed the quantum (gauge) reduced full Hamiltonian constraint, including both the Euclidean and the Lorentzian terms, and computed its expectation value on the coherent states implementing the symmetry reduction---as shown in \cite{Makinen:2020rda}, we now know that the expectation value of the quantum reduced Hamiltonian corresponds to the leading order term (in the basis states' large spin expansion) of the expectation value of the full theory Hamiltonian constraint operator on the same coherent states. Descending from the full theory, the effective Hamiltonian derived in  \cite{Alesci:2018loi} is well motivated and differs drastically from all previously postulated expressions that are based on the minisuperspace quantization models (we will come back to this comparison at the end of this section). However, ambiguities plaguing the full theory construction percolate to the quantum reduced version as well. In particular, the following choices of regularization have been made in the construction of \cite{Alesci:2018loi}:
\begin{itemize}
\item[-] The non-graph-changing version of Thiemann's
regularization \cite{Thiemann:1996aw} was considered, with loop holonomy operators entering the Euclidean term  adapted to the faces of the  cuboidal graph used to construct the reduced kinematical Hilbert space.
\item[-] The graph was kept fixed, with no sum over the number of plaquettes.
\item[-] The operator was taken in the spin 1/2 fundamental representation.
\item[-] The Lorentzian term was quantized by using its expression in terms of the 3D Ricci scalar, which is a function depending solely on the fluxes and their first and  second partial derivatives, and by relying on the diagonal action of the reduced flux operators to compute its action in a straightforward manner and without ordering ambiguities.
\end{itemize}
With these general comments in mind, let us now review how the effective Hamiltonian in \cite{Alesci:2018loi} was obtained.
We will not go through the construction of the quantum reduced Hilbert space in detail (we refer the interested reader to \cite{Alesci:2018loi}  for that), but we will focus mainly on the introduction of the coherent states and the derivation of the associated leading order corrections to the effective Hamiltonian constraint which were neglected in \cite{Alesci:2018loi} and play an important role in the analysis of Section \ref{sec:dS}.

\subsection{Coherent states}

For an arbitrary spacetime $M \simeq \mathbb{R} \times \Sigma$ with $\Sigma \simeq \mathbb{R} \times S^2$ and assuming spherical symmetry, we can introduce a local set of coordinates $(t,r,\theta,\varphi)$, with $t,r\in(-\infty,\infty), \theta\in[0,\pi] ,  \varphi \in[0,2\pi]$, and write the spacetime metric as
\be\la{dsspher}
g_{ab}dx^a dx^b= -\tilde N^2 dt^2 +\tilde\Lambda^2 \big(dr+\tilde{N}^r dt\big)^2+\tilde R^2 \big(d\theta^2+\sin^2{\theta} \ d\varphi^2 \big)\,,
\ee
with $\tilde N, \tilde{N}^r, \tilde R, \tilde \Lambda$  being a priori functions of $r$ and $t$.\footnote{We use tilde to differentiate between the metric functions for the interior homogenous foliation in \eqref{4m} and the general foliation for both the exterior and interior regions in \eqref{dsspher}, as these are in general different.}
In order to perform the quantum reduction starting from the full LQG kinematical Hilbert space, a choice of spatial manifold triangulation is introduced selecting  a subclass of cuboidal graphs, where at each vertex two pairs of links are aligned to the angular directions on the 2-sphere and one pair to the radial direction (see Fig. \ref{6valvert}). We denote  the coordinate lengths of the links tangential to these three directions by $\epsilon_\theta, \epsilon_\varphi$, and $\epsilon_r$.

 \begin{figure}[h]
 \includegraphics[width=6cm]{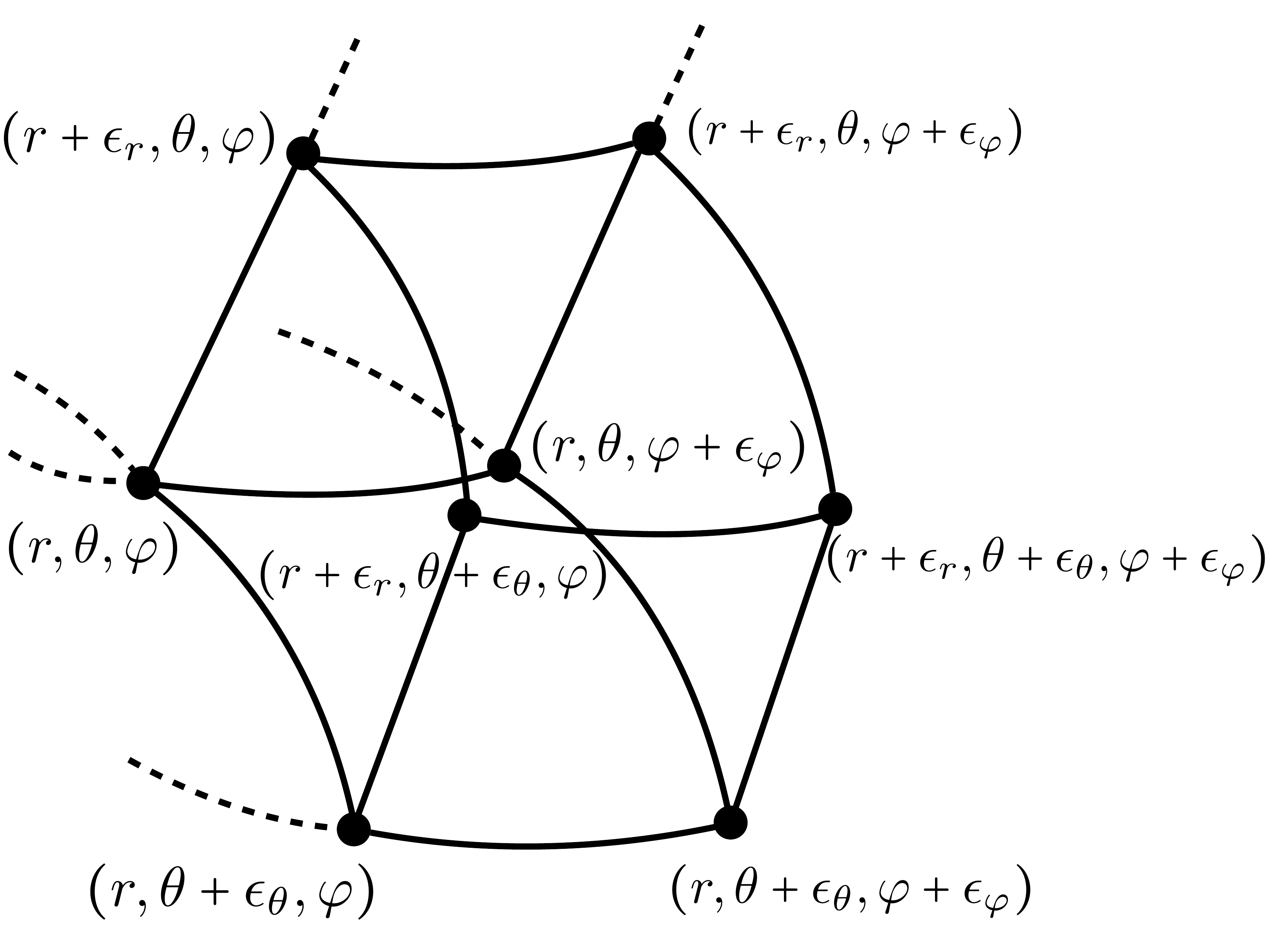}
\caption{Cuboidal graph with links adapted to the local set of coordinates.}
\la{6valvert}
\end{figure}  

The Ashtekar--Barbero connection and the densitized triad variables for a spherically symmetric geometry can be written as \footnote{ The $\tau^i$ represents an anti-Hermitian basis in the $\su(2)$ internal space, with $[\tau_i,\tau_j]=\epsilon_{ij}\!^k\tau_k$.}
\ba \label{ash}
E&=& E^r(t,r)\sin{\theta} \tau_3\partial_r+E^1(t,r)\tau_1 \sin{\theta}\partial_\theta+E^1(t,r)\tau_2 \partial_\varphi \,,\la{SE2}\\
A&=& A_r(t,r)\tau_3 dr+\left[A_1(t,r)\tau_1 + A_2(t,r)\tau_2\right]d\theta + \sin{\theta} \left[A_1(t,r)\tau_2 -A_2(t,r)\tau_1\right]d\varphi
+\cos{\theta}\tau_3 d\varphi\,.\la{Sconn2}
\ea
This represents the classical data around which we want to peak the quantum states in order to implement the symmetry reduction.
More precisely, following the construction of \cite{Thiemann:2000ca}, we introduced in \cite{Alesci:2018loi}  the  quantum reduced coherent states in the compact notation
\be\la{co-st}
\psi^{\delta_\ell}_{G}(g_\ell)= \sum_{j_\ell=0}^\infty\sum_{\bar m_\ell, \bar n_\ell=\pm j_\ell}(2j_\ell+1)(\psi^{\delta_\ell}_{G})^{j_\ell}_{ \bar n_\ell \bar m_\ell} \,{}^\ell\!D^{j_\ell}_{\bar m_\ell \bar n_\ell}(g_\ell^{-1})\,,
\ee
with the matrix coefficients $(\psi^\delta_{G})^{j_\ell}_{ \bar n_\ell \bar m_\ell} $ explicitly given by 
\begin{subequations}
\ba
\psi^{\delta_r}_{G}(g_r)
&=&\sum_{j_r=0}^\infty \sum_{\bar m_r}(2j_r+1)e^{-\frac{{\delta_r}}{2}j_r(j_r+1)} e^{{\delta_r}\bar m_r \frac{ \Delta^2_r  E^{r}\sin{\theta}}{\kappa\gamma \ell_{\va P}^2}}
D^{j_r}_{\bar n_r \bar m_r }(e^{\epsilon_r A_r\tau_3}) \,D^{j_r}_{\bar m_r \bar n_r}(g_r^{-1})\,,\la{psir}\\
\psi^{\delta_\theta}_{G}(g_\theta)
&=& \sum_{j_\theta=0}^\infty\sum_{\bar m_\theta, \bar n_\theta}(2j_\theta+1)e^{-\frac{{\delta_\theta}}{2}j_\theta(j_\theta+1)} 
 e^{{\delta_\theta}\bar m_\theta \frac{ \Delta^2_\theta E^1 \sin{\theta}}{\kappa\gamma \ell_{\va P}^2 }}
{}^x\!D^{j_\theta}_{\bar n_\theta \bar m_\theta}\left(e^{\epsilon_\theta(A_1\tau_1 + A_2\tau_2)} \right)
 {}^x\!D^{j_\theta}_{\bar m_\theta \bar n_\theta}(g_\theta^{-1})
 \,,\la{psit}\\
\psi^{\delta_\varphi}_{G}(g_\varphi)
&=& \sum_{j_\varphi=0}^\infty\sum_{\bar m_\varphi, \bar n_\varphi}(2j_\varphi+1)e^{-\frac{{\delta_\varphi}}{2}j_\varphi(j_\varphi+1)} 
e^{{\delta_\varphi}\bar m_\varphi\frac{ \Delta^2_\varphi E^1 }{\kappa\gamma \ell_{\va P}^2}}
{}^y\!D^{j_\varphi}_{\bar n_\varphi \bar m_\varphi}\left(e^{\epsilon_\varphi\left((A_1\tau_2 -A_2\tau_1)\sin{\theta}
\right)} \right)
{}^y\!D^{j_\varphi}_{\bar m_\varphi \bar n_\varphi}(g_\varphi^{-1}) 
\,,
\ea\la{psiv}
\end{subequations}
where  $\ell_{\va P}$ is the Planck length, $\delta_\ell\geq 0$ are dimensionless spread parameters governing the semi-classicality of the states and $\delta_\ell\rightarrow0$ in the classical limit.
The notation ${}^\ell\!D^{j_\ell}_{\bar m_\ell \bar n_\ell}(g_\ell)$ is used to indicate the Wigner matrix elements in the  $j_\ell$-spin representation for the $\SU(2)$ group element $g_\ell$ corresponding to the holonomy along the link in the $\ell$-direction of the  local tangent space, with basis states adapted to the local coordinate system of the metric \eqref{dsspher} \footnote{In \cite{Alesci:2018loi} the tangent direction $r$ was aligned to the internal direction $z$, while the tangent angular directions on the 2-sphere $(\theta, \varphi)$ and the internal directions $(x,y)$ had a relative mismatch reflecting   a   residual $\U(1)$ gauge symmetry. Since here we are only interested in the expectation value of the gauge invariant Hamiltonian constraint, we can set this $\U(1)$ angle to zero and consider $(\theta, \varphi)$ as aligned to $(x,y)$ without loss of generality. }. The magnetic numbers are such that $\bar m_\ell, \bar n_\ell=\pm j_\ell$.
We can define the normalized quantum reduced coherent states as
\be\la{co-st-norm}
\widetilde{ \psi^{\delta_\ell}_{G}}(g_\ell)=\frac{\psi^{\delta_\ell}_{G}(g_\ell)}{|\psi^{\delta_\ell}_{G}(g_\ell)|}  \,.
\ee

The coefficients $\psi^{\delta_\ell}_{G}(j_\ell) $ in the coherent states  are Gaussian weights peaked around the semiclassical values
$\widetilde j_\ell=\Delta^2_\ell j^0_\ell $, with $j^0_\ell$ given by
\begin{subequations}\la{j0}
\ba
j^{\va 0}_r&=&
\frac{E^r \sin{\theta}}{\kappa \ell_{\va P}^2\gamma}
\,,\\
j^{\va 0}_\theta&=&
\frac{E^1  \sin{\theta}}{\kappa \ell_{\va P}^2\gamma}
\,,\\
j^{\va 0}_\varphi&=&
\frac{E^1}{\kappa \ell_{\va P}^2\gamma}
\,,
\ea
\end{subequations}
 and $\Delta^2_\theta=\epsilon_r \epsilon_\varphi, \Delta^2_\varphi=\epsilon_r \epsilon_\theta, \Delta^2_r=\epsilon_\theta \epsilon_\varphi$, with $\kappa=8\pi$.
In the limit where $\widetilde j_\ell\gg1$, 
the expectation value of a function of  the reduced flux operators on the normalized quantum reduced coherent states yields the function evaluated on the classical data $\widetilde j_\ell$ plus coherent state corrections. The lowest order corrections are given by
\be\la{f-corr}
\langle \widetilde{ \psi^{\delta_\ell}}| f(\hat E) |\widetilde{ \psi^{\delta_\ell}} \rangle =f(\widetilde j_\ell) + \frac{\partial^2_{j_\ell} f(\widetilde j_\ell)}{2\delta_\ell}+\mathcal O(\delta_\ell^{-2} )\,.
\ee
 
  Let us analyze the corrections to the effective Hamiltonian induced by the use of coherent states peaked around the classical data to compute the expectation value of the Hamiltonian constraint. We consider also the inverse volume corrections, as a priori these two kinds of corrections can be of the same order (although we will later see that the inverse volume ones are in fact sub-dominant). Both corrections manifest themselves in the flux dependent piece of the Euclidean term in the Hamiltonian constraint in the connection formulation. A priori, coherent state corrections would appear also in the Lorentzian  term, that was regularized in  \cite{Alesci:2018loi} in terms of the densitized triad variables and their derivatives via the Ricci scalar expression.  However, as it will be shown below, the leading term of the Lorentzian piece of the effective Hamiltonian is already sub-leading with respect to the Euclidean piece, which justifies our neglecting of the coherent state corrections from this contribution for the level of approximation we are considering here.  Let us thus explicitly list both the coherent state and inverse volume corrections in the effective Euclidean  Hamiltonian coming from the expectation value of the flux dependent piece, according to \eqref{f-corr}. Given the regularization scheme adopted in  \cite{Alesci:2018loi}, we need to consider the following three contributions to the Euclidean constraint per vertex:
 \begin{subequations}\la{corrections}
\ba
 \sum_{\mu=\pm1/2}s(\mu)\sqrt{j_r j_\varphi(j_\theta+\mu)}&\approx&\frac{\tilde j_r\tilde  j_\varphi}{2\sqrt{\tilde j_r\tilde  j_\varphi \tilde  j_\theta} }
 \left(1+\frac{1}{32 \tilde  j_\theta^2}
 -\frac{1}{8 \delta_r  \tilde  j_r^2}
 +\frac{3}{8 \delta_\theta  \tilde  j_\theta^2}
 -\frac{1}{8 \delta_\varphi  \tilde  j_\varphi^2}
 \right)\,,\\
 \sum_{\mu=\pm1/2}s(\mu)\sqrt{j_r j_\theta(j_\varphi+\mu)}&\approx&\frac{\tilde j_r\tilde  j_\theta}{2\sqrt{\tilde j_r\tilde  j_\varphi \tilde  j_\theta} }
 \left(1+\frac{1}{32 \tilde  j_\varphi^2}
 -\frac{1}{8 \delta_r  \tilde  j_r^2}
 -\frac{1}{8 \delta_\theta  \tilde  j_\theta^2}
 +\frac{3}{8 \delta_\varphi  \tilde  j_\varphi^2}
 \right)\,,\\
 \sum_{\mu=\pm1/2}s(\mu)\sqrt{j_\theta j_\varphi(j_r+\mu)}&\approx&\frac{\tilde j_r\tilde  j_\theta}{2\sqrt{\tilde j_r\tilde  j_\varphi \tilde  j_\theta} }
 \left(1+\frac{1}{32 \tilde  j_r^2}
 +\frac{3}{8 \delta_r  \tilde  j_r^2}
 -\frac{1}{8 \delta_\theta  \tilde  j_\theta^2}
  -\frac{1}{8 \delta_\varphi  \tilde  j_\varphi^2}
 \right)\,,
\ea
\end{subequations} 
where $ \delta_r, \delta_\theta, \delta_\varphi$  are free dimensionless parameters at this stage.

\subsection{Effective Hamiltonian}

In order to simplify the quasi-local expression for the full effective Hamiltonian  derived in  \cite{Alesci:2018loi}, a sum over the angular plaquettes can be performed to integrate out the fluctuations around the spherical symmetry of the effective solution we aim to arrive at. Let us stress that this process of integrating out some degrees of freedom coming from the full theory structure is crucially different from freezing out these degrees of freedom from the get-go as done in the reduced quantization models. In fact, even after the sum over the angular plaquettes is performed, an imprint of the graph structure along any given 2-sphere foliating the spatial leaves remains in the resulting effective Hamiltonian. This will introduce modifications with respect to the Hamiltonians postulated in minisuperspace models which have drastic implications for the effective dynamics, as elucidated in \cite{Alesci:2019pbs}.

The sum over a given 2-sphere plaquette can be approximated  as
\ba\la{sum}
\lim_{\epsilon_\theta, \epsilon_\varphi \to 0}\sum_{p\in S^2}&=&
\lim_{\epsilon_\theta, \epsilon_\varphi \to 0}\frac{1}{\epsilon_\theta\epsilon_\varphi}
\int_0^{2\pi} d\varphi \int_0^\pi d\theta\,.
\ea

The quasi-local expression for the total  effective Hamiltonian constraint at a given vertex was derived in  \cite{Alesci:2018loi}. If we use the approximation \eqref{sum} to perform the sum over the angular plaquettes and include both the inverse volume and the coherent state corrections obtained in the previous section, the final expression for the expectation value of the  total Hamiltonian constraint operator reads (we use the superscripts IV and CS to stress the inclusion of  inverse volume and coherent state corrections with respect to our previous analyses):
\ba
&& -\frac{2\kappa \gamma^2}{2 \mathfrak{L}_0} \lim_{\epsilon_\theta, \epsilon_\varphi \to 0} \sum_{p\in S^2}\mathcal H^{\rm \va IV+CS}\n\\
&&=\lim_{\epsilon_\theta, \epsilon_\varphi\to 0}\Bigg\{
{\color{black}
\left(
1
+\frac{(\kappa\ell_{\va P}^2\gamma)^2}{32(    \epsilon_r \epsilon_\varphi E^1)^2}\frac{ \epsilon_\theta  \epsilon_\varphi  }{\pi^2}
-\frac{(\kappa\ell_{\va P}^2\gamma)^2}{8\delta_r(    \epsilon_\theta \epsilon_\varphi E^r)^2}
+\frac{3(\kappa\ell_{\va P}^2\gamma)^2}{8\delta_\theta (   \epsilon_r  \epsilon_\varphi E^1)^2}
-\frac{(\kappa\ell_{\va P}^2\gamma)^2}{8\delta_\theta (   \epsilon_r  \epsilon_\varphi E^1)^2}
\right)
} \n\\
&&\times
\frac{4 \pi^2}{\epsilon_\varphi } \sqrt{E^r}
 \Bigg[
\Bigg(h_0 \left[\left(\sqrt{A_1^2(r+\epsilon_r)+A_2^2(r+\epsilon_r)} +\sqrt{A_1^2(r)+A_2^2(r)}\right)\frac{\epsilon_\varphi}{2}\right]\n\\
&& + h_0 \left[\left(\sqrt{A_1^2(r+\epsilon_r)+A_2^2(r+\epsilon_r)} -\sqrt{A_1^2(r)+A_2^2(r)}\right)\frac{\epsilon_\varphi}{2}\right]\Bigg)
\n\\
&&\times
\frac{\left(
 \sin{\left[\frac{A_r(r)+A_r(r+\epsilon_r)}{2}\epsilon_r\right]}A_1(r+\epsilon_r)
+\cos{\left[\frac{A_r(r)+A_r(r+\epsilon_r)}{2}\epsilon_r\right]}A_2(r+\epsilon_r)
\right)}{\sqrt{A_1^2(r+\epsilon_r)+A_2^2(r+\epsilon_r)}}
\n\\
&&-
\Bigg(h_0 \left[\left(\sqrt{A_1^2(r+\epsilon_r)+A_2^2(r+\epsilon_r)} +\sqrt{A_1^2(r)+A_2^2(r)}\right)\frac{\epsilon_\varphi}{2}\right]\n\\
&& - h_0 \left[\left(\sqrt{A_1^2(r+\epsilon_r)+A_2^2(r+\epsilon_r)} -\sqrt{A_1^2(r)+A_2^2(r)}\right)\frac{\epsilon_\varphi}{2}\right]\Bigg)
\n\\
&&\times
\frac{\left(\sin{\left[ \frac{A_r(r+\epsilon_r) - A_r(r)}{2}\epsilon_r\right]}A_1(r)+\cos{\left[ \frac{A_r(r+\epsilon_r) - A_r(r)}{2}\epsilon_r\right]}A_2(r) \right)
}{\sqrt{A_1^2(r)+A_2^2(r)}}
\Bigg]\n\\
&&+
{\color{black}
\left(
1
+\frac{(\kappa\ell_{\va P}^2 \gamma)^2}{32(    \epsilon_r \epsilon_\theta E^1)^2}\frac{ \epsilon_\theta  \epsilon_\varphi  }{2\pi^2}
-\frac{(\kappa\ell_{\va P}^2\gamma)^2}{8\delta_r(    \epsilon_\theta \epsilon_\varphi E^r)^2}
-\frac{(\kappa\ell_{\va P}^2\gamma)^2}{8\delta_\theta (   \epsilon_r  \epsilon_\varphi E^1)^2}
+\frac{3(\kappa\ell_{\va P}^2\gamma)^2}{8\delta_\varphi (   \epsilon_r  \epsilon_\theta E^1)^2}
\right)
} \n\\
&&\times
\frac{16 \pi}{ \epsilon_\theta} \sqrt{E^r}
\Bigg[
\cos{\left[ \frac{\sqrt{A_1^2(r)+A_2^2(r)}}{2}\epsilon_\theta \right]}
\sin{\left[\frac{\sqrt{A_1^2(r+\epsilon_r)+A_2^2(r+\epsilon_r)}}{2}\epsilon_\theta\right]}
\n\\
&&\times
\frac{\left(
 \sin{\left[\frac{A_r(r)+A_r(r+\epsilon_r)}{2}\epsilon_r\right]}A_1(r+\epsilon_r)
+\cos{\left[\frac{A_r(r)+A_r(r+\epsilon_r)}{2}\epsilon_r\right]}A_2(r+\epsilon_r)
\right)}
{\sqrt{A_1^2(r+\epsilon_r)+A_2^2(r+\epsilon_r)}}
\n\\
&&-
\sin{\left[ \frac{\sqrt{A_1^2(r)+A_2^2(r)}}{2}\epsilon_\theta  \right]}
\cos{\left[\frac{\sqrt{A_1^2(r+\epsilon_r)+A_2^2(r+\epsilon_r)}}{2}\epsilon_\theta\right]}\n\\
&&\times
\frac{\left(\sin{\left[ \frac{A_r(r+\epsilon_r) - A_r(r)}{2}\epsilon_r\right]}A_1(r)+\cos{\left[ \frac{A_r(r+\epsilon_r) - A_r(r)}{2}\epsilon_r\right]}A_2(r) \right) }
{\sqrt{A_1^2(r)+A_2^2(r)}}
\Bigg]\n\\
&&+
{\color{black}
\left(
1
+\frac{(\kappa \ell_{\va P}^2\gamma)^2}{32(   \epsilon_\theta \epsilon_\varphi E^r )^2}\frac{ \epsilon_\theta  \epsilon_\varphi  }{\pi^2}
+\frac{3(\kappa\ell_{\va P}^2\gamma)^2}{8\delta_r(    \epsilon_\theta \epsilon_\varphi E^r)^2}
-\frac{(\kappa\ell_{\va P}^2\gamma)^2}{8\delta_\theta (   \epsilon_r  \epsilon_\varphi E^1)^2}
-\frac{(\kappa\ell_{\va P}^2\gamma)^2}{8\delta_\varphi (   \epsilon_r  \epsilon_\theta E^1)^2}
\right)
} \n\\
&&\times
\frac{4 \pi^2 \epsilon_r} {\epsilon_\theta \epsilon_\varphi}\frac{E^1}{\sqrt{E^r}}
\sin{\left[\sqrt{A_1^2(r)+A_2^2(r)}\epsilon_\theta\right]}
h_0\left[\sqrt{A_1^2(r)+A_2^2(r)}\epsilon_\varphi \right]
\n\\
&&+\frac{32 \pi \gamma^2\epsilon_r}{\epsilon_\theta^2} \frac{  E^1(r)}{ \sqrt{E^r(r)}}
     \cos{\left[\epsilon_\theta\right]} \left(\sin{\left[\frac{\epsilon_\theta}{2 }\right]}\right)^2\n\\
&&- (1+\gamma^2) \frac{2\pi}{\epsilon_r} \frac{1}{ \sqrt{E^r(r)} \big(E^1(r)\big)^2} \n\\
&\times & \Bigg[E^1(r) \Big(    \left[E^r(r+\epsilon_r) -E^r(r)\right]^2 
+  4  E^r(r)  \left[ E^r(r+2\epsilon_r) -2E^r(r+\epsilon_r)+ E^r(r)\right] \Big) \n\\
&&- 4  E^r (r)\left [E^r (r+\epsilon_r)-E^r (r)\right] \left[E^1(r+\epsilon_r)-E^1(r)\right] 
\Bigg]\Bigg\}\,,\la{Heff-int}
\ea
where we see the appearance of the Struve function of zeroth order  $h_0[x]$ as mentioned in the introduction section. We come back to this feature in Section \ref{sec:Heff-int}.

 \section{Effective Hamiltonian in the interior region}\la{sec:interior}

Let us now show how the lengthy expression  for the effective Hamiltonian in Eq. \eqref{Heff-int}, valid for both the exterior and interior regions of the black hole, assumes a much simpler form when adapted to a homogeneous foliation as in the interior region. 
To this end, let us set $N^r=0$ and replace the coordinates $(t,r)$ with $(\tau,x)$ of the interior metric \eqref{4m}.

\subsection{Choice of coherent state parameters}\la{sec:coherent-state}

Our predictions for the black hole interior
region are state-dependent. This is not a novel feature of our construction, rather it is an overall ambiguity that is present in all previous minisuperspace models as well. In the LQC framework, the full theory graph structure is absent and such state-dependence is interpreted a posteriori as a choice of regularization scheme  (see, e.g., \cite{Modesto:2005zm,Ashtekar:2005qt,  Bohmer:2007wi, Chiou:2008nm, Brannlund:2008iw, Ashtekar:2018lag}). This choice is related to different embeddings of a fiducial discrete structure in the effective theory and is aimed at giving a physical interpretation to the polymer parameters introduced as cut-off. There is a certain level of  ambiguity in this prescription  which allows for a wide spectrum of regularization schemes. The choice  can  be justified only a posteriori, by verifying that the  effective solution satisfies some desired reasonable physical demands.
On the other hand, having the full theory structure to begin with, our choice of regularization scheme determining the  quantum states that we attribute to the black hole interior region can be guided by clear geometrical considerations. We now explain how our construction of quantum states is related to the discrete geometrical information  associated  to our choice of graph structure in a clear-cut way, reducing considerably the arbitrariness inherent to minisuperspace models. 

\subsubsection{Quantum parameters}
To begin, recall that the fundamental building blocks 
of our graph triangulating the leaves of foliation  are cuboidal cells that are formed by eight six-valent vertices. Each vertex lies on a given 2-sphere, with
 two links tangent to the angular directions $\theta$ and $ \varphi$  and one along the orthogonal $x$ direction. 
 We denote the coordinate length for the 2-sphere tangent links by  $\epsilon_\theta= \epsilon_\varphi:= \epsilon$ and  the coordinate length for  the orthogonal link by $\epsilon_x$. We can define these coordinate lengths as
\be\la{Ne}
\epsilon :=\frac{2\pi}{\mathcal N}\,,\quad \quad\epsilon_x :=\frac{\mathfrak{L}_0}{\mathcal  N_x}\,,
\ee
where $\mathcal N$ and $ \mathcal N_x$ are two integers such that $\mathcal  N^2/2$ is the total number of plaquettes on the 2-sphere\footnote{The factor 1/2 comes from the fact that the coordinate lengths for the links along the two angular coordinates $\theta$ and $ \varphi$ can be written as $\epsilon_\theta=\pi/\mathcal N_\theta, \epsilon_\varphi=2\pi/\mathcal N_\varphi$. Requiring $\epsilon_\theta=\epsilon_\varphi=\epsilon$ implies $2\mathcal N_\theta=\mathcal N_\varphi=\mathcal N$, so that the total number of plaquettes covering the 2-sphere is given by $\mathcal N_\theta \mathcal N_\varphi=\mathcal N^2/2$.} and $\mathcal N_x$ is the total number of plaquettes in the $x$-direction for a given fiducial length $\mathfrak{L}_0$. As we will illustrate more in detail below, in order to approximate an integral over the three spatial directions with a sum over plaquettes (and vice-versa), we need to consider the limit where $\epsilon, \epsilon_x \ll1 $, or equivalently $\mathcal  N,\mathcal N_x\gg1$. In this limit, we can express the area of a given 2-sphere $S$ in the interior region as
\be\la{ep1}
A(R)= 4 \pi R^2 
=8\pi \gamma\ell^2_{\va P}  \sum_{p\in S}  \tilde j^p_x 
\simeq 4\pi \gamma\ell^2_{\va P}  j_x\mathcal N^2\,,
\ee
where the sum is over all plaquettes that tessellate  the given  2-sphere with radius $R$, and $ \tilde j_x^p$ is the spin number associated with the link dual to the given plaquette $p$ in the coherent state.  In the limit $\mathcal  N\gg 1$, we  approximate this sum with the product of a single (average) spin number $j_x$ times  the total number of plaquettes on $S$. Similarly, we can express the volume of a given spatial hypersurface $\Sigma$ as (recall that we are using $\mathfrak{L}_0$ as a regulator for integrals over the $x$ direction)\footnote{The spectrum of the volume operator can be easily computed in the quantum reduced loop gravity framework where the reduced flux operators become diagonal  and the contribution at each node is given by the sum of the contributions from all cubes around the given vertex \cite{Alesci:2018loi}. The cuboidal graph structure adopted in the quantum reduction yields four times the 3-valent vertex eigenvalue.}
\be\la{ep2}
V(\Sigma)=8\pi \mathfrak{L}_0 \sqrt{E^x(\tau)} E^1(\tau)\simeq
4 (8\pi\gamma \ell^2_{\va P})^{3/2}  j\sqrt{j_x} \mathcal N_x \mathcal N^2 \,,
\ee
where we have denoted the average spin number associated with the links dual to the plaquettes in both  $(\theta, x)$ and $(\varphi, x)$ planes by $j$.
From Eqs. \eqref{ep1} and \eqref{ep2} we arrive at
\ba \label{epsilons}
\epsilon&=&\frac{\alpha}{R}\,,\quad\quad \alpha:=2\pi\sqrt{ \gamma j_x}\,  \ell_{\va P}\,,\n\\
\epsilon_x  &=& \frac{\beta}{ \Lambda}\,,\quad \quad\beta:=\frac{4 \sqrt{8 \pi \gamma} \, j \,\ell_{\va P}}{ \sqrt{j_x }}\,.
\ea
One can see that the two quantum parameters are functions of metric, the Barbero--Immirzi parameter, the Planck length, and the two spin numbers that enter the definition of our coherent states.

\subsubsection{Spins}

Although our focus in this paper is on the interior geometry, it turns out that we can further constrain the class of coherent states by extending our geometrical analysis to the black hole exterior region.
The transition from the interior to the exterior region is marked by $\partial_x$ changing from being space-like to time-like.
This implies that, for an outside observer,  the spin number $j$ is associated with a generator of rotations in the time-$\theta$ or time-$\varphi$ plane, {\it i.e.} to a boost generator. This signature change for the  intrinsic metric on constant $\tau$ surfaces is accompanied by
a change for the gauge group of internal rotations from $\SU(2)$ to $\SU(1,1)$ in the Ashtekar formulation. Therefore, if we demand that our spatial manifold triangulation remains consistent on both sides of the horizon, it is consistent to require 
\be\la{sim}
j=\gamma j_x\,,
\ee
a relation between the two spin numbers $j_x$ and $j$ that follows from the imposition of the linear simplicity constraint \cite{Perez:2012wv}.

In the context of black hole physics, this interplay between the canonical and covariant formulations of the loop quantum gravity has proven to be very important in our understanding of thermal properties of the black hole horizon \cite{Bianchi:2012ui, Pranzetti:2013lma}. As it will shown below, it also has interesting implications for the physical predictions of our model. 

\subsubsection{Spread parameters}

Finally, we have to fix the form of the spread parameters $\delta_\ell$. In order for the expansions \eqref{corrections} to be valid, we need the condition 
\be\la{deltaj}
\delta_\ell \tilde j_\ell^2\gg1
\ee
 to be satisfied. 
We  make the following rescaling  of the spread parameters \
\begin{subequations}\la{deltas}
\ba
\delta_r&=&\frac{\pi^2 \ell_{\va P}^2 R^2}{\alpha^4 (\sin{\theta})^2} \delta_x\,,\\
\delta_\theta&=&\frac{\pi^2 \ell_{\va P}^2 R^2}{\alpha^2\beta^2 (\sin{\theta})^2} \delta\,,\\
\delta_\varphi&=&\frac{\pi^2 \ell_{\va P}^2 R^2}{\alpha^2\beta^2 } \frac{\delta}{\nu}
\,,
\ea
\end{subequations}
where $\delta, \nu, \delta_x$ are the new free dimensionless parameters entering the coherent state corrections. When solving for the effective dynamics we need to make sure that the consistency condition \eqref{deltaj} is respected by the effective solutions.

\subsection{Effective Hamiltonian}\la{sec:Heff-int}

In the interior region, the Ashtekar variables  \eqref{ash} are easily related to the interior ADM variables introduced there via the following relations \footnote{In both foliations \eqref{dsspher} and \eqref{4m} the connection component $A_2$  does not enter the reduced phase space; in the former case, it is constrained to be the lapse function by the spatial diffeomorphism in the radial direction, and in the latter it simply vanishes.}
\ba
&& E^x = R^2\,,\quad\quad E^1 = R \Lambda\,, \nonumber\\
&&A_x = - \frac{\gamma G}{R} \Big(P_R - \frac{\Lambda}{R} P_\Lambda\Big)\,,\quad\quad A_1 = - \frac{\gamma G}{R} P_\Lambda\,,\quad\quad A_2 =0\,.\la{ADM-A}
\ea

It follows that, in light of \eqref{epsilons} and with our rescaling \eqref{deltas} of the spread parameters,  the  total effective Hamiltonian constraint   for the interior homogenous  foliation   (after dividing by $\epsilon_x $ as well)  is given by
 
 
 
\ba \label{heff-IV+CS}
{\mathcal H}^{\rm \va IV+CS}_{\rm int} =
&-& \frac{\mathfrak{L}_0 R^2\Lambda}{2 \gamma^2 G \beta \alpha^2}\n\\
\times &\Bigg\{&
 \alpha
{\color{black}
\left(
1
+\frac{2\ell_{\va P}^4\gamma^2}{ \beta^2 R^2 }
-\frac{8\ell_{\va P}^2\gamma^2}{\delta_x  R^2}
+\frac{8(3-\nu)\ell_{\va P}^2\gamma^2}{\delta R^2 }
\right)
}
 \sin{\left[\frac{\gamma G\beta [P_R R - P_\Lambda \Lambda]}{R^2\Lambda}\right]}
 \pi h_0\left[\frac{\gamma G \alpha P_\Lambda}{R^2}\right]
\n\\
&+&
\alpha
{\color{black}
\left(
1
+\frac{\ell_{\va P}^4\gamma^2}{ \beta^2 R^2 }
-\frac{8\ell_{\va P}^2\gamma^2}{\delta_x   R^2}
+\frac{8(3\nu-1)\ell_{\va P}^2\gamma^2}{\delta R^2 }
\right)
}
 \sin{\left[\frac{\gamma G\beta [P_R R - P_\Lambda \Lambda]}{R^2\Lambda}\right]}
2 \sin{\left[\frac{\gamma G \alpha P_\Lambda}{R^2}\right]} 
\n\\
&+&
\beta
\left(
8 \gamma^2 
 \cos{\left[\frac{\alpha}{R}\right]}  \sin{\left[\frac{\alpha}{2R}\right]}^2 +
{\color{black}
\left(
1
+\frac{2\ell_{\va P}^4\gamma^2}{\alpha^2R^2}
+\frac{24\ell_{\va P}^2\gamma^2}{\delta_x  R^2}
-\frac{8(1+\nu)\ell_{\va P}^2\gamma^2}{\delta R^2 }
\right)
}
 \pi \sin{\left[\frac{\gamma G \alpha P_\Lambda}{R^2}\right]} h_0 \left[\frac{\gamma G \alpha P_\Lambda}{R^2}\right] 
 \right) 
 \Bigg\}\,,\n\\
\ea
where the subscript ``int'' stands for ``interior''.

It should be noted that ${\mathcal H}^{\rm \va IV+CS}_{\rm int}$ significantly
differs from the effective Hamiltonian constraint of the
LQC minisuperspace quantization models. In fact, the appearance of the Struve function $h_0[x]$ in Eq. \eqref{heff-IV+CS} originates from  integrating over the angular coordinates  of  holonomies along links tangent to a given 2-sphere of the leaves of foliation and it is thus associated to degrees of freedom which are non-existent in the LQC approaches due to using point holonomies. Similarly, the two quantum parameters $\epsilon$ and $\epsilon_x$ 
correspond to the coordinate lengths of links of the cubic cells in the chosen graph which enter the definition of the coherent states \eqref{co-st}.
They can be thought of as the discretization parameters for the graphs that have been adapted to constant $\tau$ surfaces. As opposed to the previous polymer quantization models, the availability of the full theory geometrical setup allows us to determine the expression of these two crucial parameters in a straightforward and unambiguous manner, as  in \eqref{epsilons}. The second main difference induced by the inclusion of the 2-sphere degrees of freedom in our analysis is encoded in the term proportional to $\gamma^2$ in expression  \eqref{heff-IV+CS}. This contribution comes from the Lorentzian piece of the scalar Hamiltonian and only the leading term in its $\epsilon$-expansion is included in the previous proposals, while the higher order corrections are neglected. We close this section by pointing out that our effective Hamiltonian for the interior region given in Eq. \eqref{heff-IV+CS} reduces to the  minisuperspace Hamiltonian of \cite{Ashtekar:2018cay} after the following replacements 
 \be
 h_0[x]\mapsto \frac{2}{\pi} \sin{[x]}\,,\quad \quad
  \cos{[\epsilon]}  \sin{\left[\frac{\epsilon}{2}\right]}^2  \mapsto \frac{\epsilon^2}{4}\,,
 \ee
 and, of course, neglecting both inverse volume and coherent state corrections.
 


\section{Interior effective dynamics}\la{sec:eff-eq}

We now shift gears to discuss the effective Hamilton's equations. As in the classical system, we first choose a hypersurface $\Sigma$ in vicinity of the black hole's event horizon where initial data is specified \footnote{Here we are assuming that a black hole event horizon exists. Note that any analysis that is purely based on the interior geometry does not imply the existence of the event horizon. The question of whether the black hole event horizon exists will have to be determined by a more sophisticated analysis that subsumes the entire spacetime.}. First, note that there is a large amount of freedom in selecting the initial data on $\Sigma$. Indeed, should we take the extreme case where $\Sigma$ tends to the event horizon, it follows from Eq. \eqref{heff-IV+CS} that it is sufficient to require $\Lambda \Big|_{\Sigma} = P_{\Lambda} \Big|_{\Sigma} = 0$ and $0 < |R|, |P_R| < \infty$. Nevertheless, this level of arbitrariness is significantly reduced if we make use of the classical geometry in setting up the initial data, which in turn requires that $  1\ll\sqrt{j} \ll m/m_p $, where $m_p = \sqrt{\hbar c/G}$ is the Planck mass \footnote{Due to Eq. \eqref{sim}, $j$ and $j_x$ are expected to be comparable in magnitude as long as $\gamma\sim \mathcal O(1)$. We will see below how the results of our asymptotic analysis are consistent with this assumption.}. This would be a well motivated approach given that we are limiting our analysis to the interior regions of astrophysical black holes.  The latter condition on the black hole's mass translates to 
\be 
\epsilon \Big|_{\Sigma} \sim  \frac{m_p \sqrt{j}}{m} \ll 1\,, \quad\quad \epsilon_x \Big|_{\Sigma} \sim \frac{R}{\Lambda} \epsilon \Big|_{\Sigma} \ll \frac{R}{\Lambda}\Big|_{\Sigma}\,,
\ee
which are satisfied for sufficiently massive black holes in vicinity of their event horizons. Not surprisingly, expanding in powers of $\epsilon$ and $\epsilon_x$ gives 
$\mathcal{H}_{\rm c}$ as the lowest order term in $\mathcal{H}^{{\rm \va IV+CS}}_{\rm int}$. Thus, one can reliably adjust the classical data near the event horizon to incorporate $\mathcal{O}(\epsilon, {\epsilon_x})$ corrections. Bear in mind that as $\Sigma$ tends to the event horizon, the error in $\mathcal{H}^{\rm \va IV+CS}_{\rm int}(R_{\rm c}, \Lambda_{\rm c}, P_{R_{\rm c}}, P_{\Lambda_{\rm c}}) = 0$ becomes vanishingly small.

To solve the dynamical equations,  we choose to smear $\mathcal{H}^{\rm \va IV+CS}_{\rm int}$ given in Eq. \eqref{heff-IV+CS} with the following lapse function:
 \be \la{N}
N= -\frac{\gamma \epsilon R}{ G m\Big[\sin{\Big(\frac{\gamma G \epsilon P_\Lambda}{R}\Big)} + \frac{\pi}{2} h_0 \Big( \frac{\gamma G \epsilon P_\Lambda}{R}\Big)\Big]}\,.
\ee
This choice of lapse function reduces to \eqref{Nc} in the limit $\hbar \rightarrow 0$. For $m\gg m_p $,   the black hole's event horizon is still located  near  the time coordinate $\tau = 0$. Unless a second inner Killing horizon is reached, $\tau$ can be extended all the way to $- \infty$.

For later convenience, let us replace $\tau$ with the new variable  $z :=\exp{(-\tau/\ell)}$, where $\ell$ is  some length scale whose physical meaning will become clear in the following. Denoting $z$-derivatives by prime, the evolution equations for $R$ and $P_\Lambda$ read:
\ba \label{Rdot-CS}
 -\frac{z}{\ell}R' &=&
\frac{1}{2 \mathfrak{L}_0}\frac{\partial{\mathcal H}^{\rm \va IV+CS}_{\rm int} [N]}{\partial P_R}\n\\
&=&
 \frac{R}{2 G m }\cos{\left[\gamma G \beta \left(\frac{P_R }{ R \Lambda }- \frac{P_\Lambda}{R^2}\right)\right]}\n\\
& \times&\Bigg\{
{\color{black}
\left(
1
+\frac{2\ell_{\va P}^4\gamma^2}{ \beta^2 R^2 }
-\frac{8\ell_{\va P}^2\gamma^2}{\delta_x  R^2}
+\frac{8(3-\nu)\ell_{\va P}^2\gamma^2}{\delta R^2 }
\right)
}
\frac{ \pi h_0\left[\frac{\gamma G \alpha P_\Lambda}{R^2}\right]}{(2 \sin{\left[\frac{\gamma G \alpha P_\Lambda}{R^2}\right]} + \pi h_0\left[\frac{\gamma G \alpha P_\Lambda}{R^2}\right])}\n\\
&+&
{\color{black}
\left(
1
+\frac{\ell_{\va P}^4\gamma^2}{ \beta^2 R^2 }
-\frac{8\ell_{\va P}^2\gamma^2}{\delta_x   R^2}
+\frac{8(3\nu-1)\ell_{\va P}^2\gamma^2}{\delta R^2 }
\right)
}
\frac{ 2 \sin{\left[\frac{\gamma G \alpha P_\Lambda}{R^2}\right]
}}{(2 \sin{\left[\frac{\gamma G \alpha P_\Lambda}{R^2}\right]} + \pi h_0\left[\frac{\gamma G \alpha P_\Lambda}{R^2}\right])}
\Bigg\}
\,, \\
 -\frac{z}{\ell} {P}'_\Lambda&=&
-  \frac{1}{2 \mathfrak{L}_0}\frac{\partial {\mathcal H}^{\rm \va IV+CS}_{\rm int}[N]}{\partial \Lambda}\n\\
&\hat{=}&\frac{R P_R}{2 Gm\Lambda}\cos{\left[\gamma G \beta \left(\frac{P_R }{ R \Lambda }- \frac{P_\Lambda}{R^2}\right)\right]}\n\\
& \times&\Bigg\{
{\color{black}
\left(
1
+\frac{2\ell_{\va P}^4\gamma^2}{ \beta^2 R^2 }
-\frac{8\ell_{\va P}^2\gamma^2}{\delta_x  R^2}
+\frac{8(3-\nu)\ell_{\va P}^2\gamma^2}{\delta R^2 }
\right)
}
\frac{ \pi h_0\left[\frac{\gamma G \alpha P_\Lambda}{R^2}\right]}{(2 \sin{\left[\frac{\gamma G \alpha P_\Lambda}{R^2}\right]} + \pi h_0\left[\frac{\gamma G \alpha P_\Lambda}{R^2}\right])}\n\\
&+&
{\color{black}
\left(
1
+\frac{\ell_{\va P}^4\gamma^2}{ \beta^2 R^2 }
-\frac{8\ell_{\va P}^2\gamma^2}{\delta_x   R^2}
+\frac{8(3\nu-1)\ell_{\va P}^2\gamma^2}{\delta R^2 }
\right)
}
\frac{ 2 \sin{\left[\frac{\gamma G \alpha P_\Lambda}{R^2}\right]
}}{(2 \sin{\left[\frac{\gamma G \alpha P_\Lambda}{R^2}\right]} + \pi h_0\left[\frac{\gamma G \alpha P_\Lambda}{R^2}\right])}
\Bigg\}
\,, 
\ea
where $\hat{=}$ indicates that the equation has been evaluated on-shell ({\it i.e.} $\mathcal{H}^{\rm \va IV+CS}_{\rm int}=0$ is imposed). The above equations lead to the following equation for $P_\Lambda$ which will be useful in the subsequent sections:
\ba\la{PLdot-CS}
 {P}'_\Lambda&=&\frac{R' P_R}{\Lambda}
\n\\
 &=&\frac{R' P_\Lambda}{R}\n\\
& +&
 \frac{R R'}{G\gamma\beta}
 \arccos \Bigg[    -2 G m \frac{ z}{\ell}\frac{R'}{R} \n\\
 &\times&
 \Bigg( \frac{
 \left(2 \sin{\left[\frac{\gamma G \alpha P_\Lambda}{R^2}\right]} + \pi h_0\left[\frac{\gamma G \alpha P_\Lambda}{R^2}\right]\right)
 }
 {
{
{\color{black}
\left(
1
+\frac{2\ell_{\va P}^4\gamma^2}{ \beta^2 R^2 }
-\frac{8\ell_{\va P}^2\gamma^2}{\delta_x  R^2}
+\frac{8(3-\nu)\ell_{\va P}^2\gamma^2}{\delta R^2 }
\right)
}
 \pi h_0\left[\frac{\gamma G \alpha P_\Lambda}{R^2}\right]}
+
{ 
{\color{black}
\left(
1
+\frac{\ell_{\va P}^4\gamma^2}{ \beta^2 R^2 }
-\frac{8\ell_{\va P}^2\gamma^2}{\delta_x   R^2}
+\frac{8(3\nu-1)\ell_{\va P}^2\gamma^2}{\delta R^2 }
\right)
}
2 \sin{\left[\frac{\gamma G \alpha P_\Lambda}{R^2}\right]
}}
}\Bigg)
  \Bigg]
  \,.
\ea

Finally, the evolution equations for $\Lambda$ and $P_R$ are given by
\ba\la{Ldot-CS}
-2G m\frac{z}{\ell}\frac{\Lambda'} {\Lambda}
&=&\frac{G m}{\mathfrak{L}_0 {\Lambda}}\frac{\partial {\mathcal H}^{\rm \va IV+CS}_{\rm int}[N]}{\partial P_\Lambda}\n\\
&=&
-
\cos{\left[\gamma G \beta \left(\frac{P_R }{ R \Lambda }- \frac{P_\Lambda}{R^2}\right)\right]}\n\\
& \times&\Bigg\{
{\color{black}
\left(
1
+\frac{2\ell_{\va P}^4\gamma^2}{ \beta^2 R^2 }
-\frac{8\ell_{\va P}^2\gamma^2}{\delta_x  R^2}
+\frac{8(3-\nu)\ell_{\va P}^2\gamma^2}{\delta R^2 }
\right)
}
\frac{ \pi h_0\left[\frac{\gamma G \alpha P_\Lambda}{R^2}\right]}{(2 \sin{\left[\frac{\gamma G \alpha P_\Lambda}{R^2}\right]} + \pi h_0\left[\frac{\gamma G \alpha P_\Lambda}{R^2}\right])}\n\\
&+&{\color{black}
\left(
1
+\frac{\ell_{\va P}^4\gamma^2}{ \beta^2 R^2 }
-\frac{8\ell_{\va P}^2\gamma^2}{\delta_x  R^2}
+\frac{8(3\nu-1)\ell_{\va P}^2\gamma^2}{\delta  R^2 }
\right)
}
\frac{ 2 \sin{\left[\frac{\gamma G \alpha P_\Lambda}{R^2}\right]
}}{(2 \sin{\left[\frac{\gamma G \alpha P_\Lambda}{R^2}\right]} + \pi h_0\left[\frac{\gamma G \alpha P_\Lambda}{R^2}\right])}
\Bigg\}\n\\
&+&
\frac{\pi h_{\va -1}\left[\frac{ \gamma G\alpha P_\Lambda}{R^2}\right]\left(
{\color{black}
\left(
1
+\frac{2\ell_{\va P}^4\gamma^2}{\alpha^2R^2}
+\frac{24\ell_{\va P}^2\gamma^2}{\delta_x  R^2}
-\frac{8(1+\nu)\ell_{\va P}^2\gamma^2}{\delta R^2 }
\right)
}
 2 \sin^2{\left[\frac{ \gamma G \alpha P_\Lambda}{R^2}\right]}
-8 \gamma^2 \cos{\left[\frac{\alpha}{R} \right]}  \sin^2{\left[\frac{\alpha}{2 R} \right]}
\right)}{\left( 2 \sin{\left[\frac{ \gamma G \alpha P_\Lambda}{R^2}\right]}+\pi h_0\left[\frac{ \gamma G \alpha P_\Lambda}{R^2}\right]
\right)^2}\n\\
&+&\frac{\cos{\left[\frac{ \gamma G \alpha P_\Lambda}{R^2}\right]}\left(
{\color{black}
\left(
1
+\frac{2\ell_{\va P}^4\gamma^2}{\alpha^2R^2}
+\frac{24\ell_{\va P}^2\gamma^2}{\delta_x  R^2}
-\frac{8(1+\nu)\ell_{\va P}^2\gamma^2}{\delta R^2 }
\right)
}
\pi^2 h^2_0\left[\frac{ \gamma G \alpha P_\Lambda}{R^2}\right]
-16 \gamma^2 \cos{\left[\frac{\alpha}{R} \right]}  \sin^2{\left[\frac{\alpha}{2 R} \right]}
\right)}{\left( 2 \sin{\left[\frac{ \gamma G \alpha P_\Lambda}{R^2}\right]}+\pi h_0\left[\frac{ \gamma G \alpha P_\Lambda}{R^2}\right]
\right)^2}\n\\
&+&
\frac{2 \pi \gamma^2 \alpha{\ell_P ^2}}{\beta  R^2}
{\color{black}
\left(\frac{{\ell_P ^2}}{\beta^2}+\frac{ 32(1-\nu)}{\delta}
\right)
}
\sin{\left[\gamma G \beta \left(\frac{P_R }{ R \Lambda }- \frac{P_\Lambda}{R^2}\right)\right]}
\frac{\left(
\sin{\left[\frac{ \gamma G \alpha P_\Lambda}{R^2}\right]}h_{-1}\left[\frac{ \gamma G \alpha P_\Lambda}{R^2}\right]
-
\cos{\left[\frac{ \gamma G \alpha P_\Lambda}{R^2}\right]}
h_0\left[\frac{ \gamma G \alpha P_\Lambda}{R^2}\right]
\right)
}
{\left( 2 \sin{\left[\frac{ \gamma G \alpha P_\Lambda}{R^2}\right]}+\pi h_0\left[\frac{ \gamma G \alpha P_\Lambda}{R^2}\right]
\right)^2}
\,,\n\\
\ea
\ba\la{PRdot-CS}
-\frac{z}{\ell}P_R' &=& - \frac{1}{2 \mathfrak{L}_0} \frac{\partial \mathcal{H}^{\rm \va IV+CS}_{\rm int}[N]}{\partial R}\n\\
&\hat{=}& \frac{(R P_R - 2 \Lambda P_\Lambda)}{2 G m R}
\cos{\left[\gamma G \beta \left(\frac{P_R }{ R \Lambda }- \frac{P_\Lambda}{R^2}\right)\right]}\n\\
& \times&\Bigg\{
{\color{black}
\left(
1
+\frac{2\ell_{\va P}^4\gamma^2}{ \beta^2 R^2 }
-\frac{8\ell_{\va P}^2\gamma^2}{\delta_x  R^2}
+\frac{8(3-\nu)\ell_{\va P}^2\gamma^2}{\delta R^2 }
\right)
}
\frac{ \pi h_0\left[\frac{\gamma G \alpha P_\Lambda}{R^2}\right]}{(2 \sin{\left[\frac{\gamma G \alpha P_\Lambda}{R^2}\right]} + \pi h_0\left[\frac{\gamma G \alpha P_\Lambda}{R^2}\right])}\n\\
&+&{\color{black}
\left(
1
+\frac{\ell_{\va P}^4\gamma^2}{ \beta^2 R^2 }
-\frac{8\ell_{\va P}^2\gamma^2}{\delta_x  R^2}
+\frac{8(3\nu-1)\ell_{\va P}^2\gamma^2}{\delta  R^2 }
\right)
}
\frac{ 2 \sin{\left[\frac{\gamma G \alpha P_\Lambda}{R^2}\right]
}}{(2 \sin{\left[\frac{\gamma G \alpha P_\Lambda}{R^2}\right]} + \pi h_0\left[\frac{\gamma G \alpha P_\Lambda}{R^2}\right])}
\Bigg\}\n\\
&+& \frac{\pi \Lambda P_\Lambda \sin{\Big[\frac{\alpha \gamma G P_\Lambda}{R^2}\Big]} h_{-1}\Big[\frac{\alpha \gamma G P_\Lambda}{R^2}\Big]}{G m R (2 \sin{\left[\frac{\gamma G \alpha P_\Lambda}{R^2}\right]} + \pi h_0\left[\frac{\gamma G \alpha P_\Lambda}{R^2}\right])} {\color{black}
\left(
1
+\frac{2 \ell_{\va P}^4\gamma^2}{ \alpha^2 R^2 }
+\frac{24\ell_{\va P}^2\gamma^2}{\delta_x  R^2}
-\frac{8(\nu+1)\ell_{\va P}^2\gamma^2}{\delta  R^2 }
\right)
} \n\\
&+&\frac{\pi \Lambda h_0\left[\frac{\gamma G \alpha P_\Lambda}{R^2}\right]}{G^2 m R (2 \sin{\left[\frac{\gamma G \alpha P_\Lambda}{R^2}\right]} + \pi h_0\left[\frac{\gamma G \alpha P_\Lambda}{R^2}\right])} \bigg\{\sin{\Big[\frac{\gamma G \alpha P_\Lambda}{R^2}\Big]}{\color{black}\Big(\frac{2 \ell_{\va P} ^4 \gamma}{\alpha^3}+ \frac{24 \ell_{\va P} ^2 \gamma}{\alpha \delta_x}-\frac{8 \gamma \ell_{\va P}^2 (1+\nu)}{\alpha \delta}\Big)} \n\\
&+& G P_\Lambda \cos{\Big[\frac{\gamma G \alpha P_\Lambda}{R^2}\Big]} {\color{black}\Big(1+ \frac{2 \ell_{\va P} ^4 \gamma^2}{R^2 \alpha^2}+ \frac{24 \ell_{\va P} ^2 \gamma^2}{R^2 \delta_x}-\frac{8 \gamma^2 \ell_{\va P}^2 (1+\nu)}{R^2 \delta}\Big)} \bigg\}\n\\
&+&\frac{\Lambda \sin{\left[\gamma G \beta \left(\frac{P_R }{ R \Lambda }- \frac{P_\Lambda}{R^2}\right)\right]}}{G^2 m R (2 \sin{\left[\frac{\gamma G \alpha P_\Lambda}{R^2}\right]} + \pi h_{0}\left[\frac{\gamma G \alpha P_\Lambda}{R^2}\right])} \bigg\{G \pi \frac{\alpha}{\beta} P_\Lambda h_{-1}\left[\frac{\gamma G \alpha P_\Lambda}{R^2}\right]{\color{black} \Big(1 + \frac{2 \ell_{\va P}^4 \gamma^2}{R^2 \beta^2}+\frac{8 \ell_{\va P}^2 \gamma^2 (3 - \nu)}{R^2 \delta}}\n\\
&-&{\color{black} \frac{8 \ell_{\va P}^2 \gamma^2}{R^2 \delta_x}\Big)} + 2 \pi h_{0}\left[\frac{\gamma G \alpha P_\Lambda}{R^2}\right]{\color{black}\Big(\frac{\ell_{\va P}^4 \gamma}{\beta^3}+ \frac{4 \ell_{\va P}^2 \gamma (3- \nu)}{\beta \delta}-\frac{4 \ell_{\va P}^2 \gamma}{\beta \delta_x}\Big)}\n\\
&+& \sin{\left[\frac{\gamma G \alpha P_\Lambda}{R^2}\right]} {\color{black} \Big(\frac{2 \ell_{\va P}^4 \gamma}{\beta^3}+ \frac{16 \ell_{\va P}^2 \gamma (3 \nu- 1)}{\beta \delta}-\frac{16 \ell_{\va P}^2 \gamma}{\beta \delta_x}\Big)} + \frac{2 G P_\Lambda \alpha}{\beta}\cos{\left[\frac{\gamma G \alpha P_\Lambda}{R^2}\right]} \n\\
&\times& {\color{black} \Big(1+\frac{\ell_{\va P}^4 \gamma^2}{R^2 \beta^2}+\frac{8 \ell_{\va P}^2 \gamma^2 (3 \nu -1)}{R^2 \delta} - \frac{8 \ell_{\va P}^2 \gamma^2}{R^2 \delta_x} \Big)}\bigg\}-
\frac{2\gamma\Lambda  }{G^2 m}\frac{ \left(\sin{\left[\frac{\alpha}{ R} \right]} -\sin{\left[\frac{2\alpha}{ R} \right]}\right) }
{\left( 2 \sin{\left[\frac{ \gamma G \alpha P_\Lambda}{R^2}\right]}+\pi h_0\left[\frac{ \gamma G \alpha P_\Lambda}{R^2}\right]
\right)}\,.
\ea

While solving for the effective dynamics, we replace the evolution equation for $P_R$ with the effective Hamiltonian constraint \eqref{heff-IV+CS}. This choice is justified due to the fact that the phase space is 4-dimensional; thus, any of the four evolution equations can be replaced with the more manageable constraint equation.

\section{Asymptotically de Sitter geometry for the interior}\la{sec:dS}

The Hamiltonian presented in Eq. \eqref{heff-IV+CS}
has  several free quantum parameters; $\beta$ (or $\alpha$)\footnote{Recall that we have imposed the simplicity constraint \eqref{sim} on the quantum spin numbers, which implies $\alpha/\beta = \sqrt{2 \pi}/(8 \gamma)$.}, $\gamma$, $\nu$, $\delta$, and $\delta_x$. It is therefore expected for the corresponding dynamical system to accommodate a large class of solutions. The intriguing fact particular to this Hamiltonian is that some of these geometries have asymptotic structures of special interest, and this leads us to the central question we want to address in this work. 
 Indeed, we demonstrate in this section how a judicious choice of the quantum parameters leads to an asymptotically Schwarzschild--de Sitter interior geometry as defined in \cite{Ashtekar:2014zfa}; a positive cosmological constant emerges from the quantum gravitational effects.  Somewhat of a mystery, and a surprise at the same time,  is the value for the Barbero--Immirzi parameter, $\gamma$, for this geometry which is shown to coincide with the value from the $\SU(2)$ black hole entropy calculations in LQG. We will come back to this important feature  at the end of Section \ref{sec:series}.



\subsection{Asymptotic series solution}\label{sec:series}

We aim to show that the scalar constraint equation $\mathcal{H}^{\rm \va IV+CS}_{\rm int}=0$ together with the dynamical Eqs. \eqref{Rdot-CS},  \eqref{PLdot-CS},  and \eqref{Ldot-CS} admit a solution set with the following asymptotic property: 
  \begin{subequations} \label{ansatz}
\ba
&& \lim_{z \rightarrow \infty} \Lambda(z) =\frac{1}{2} \Big(z-\frac{1}{z}\Big) + \mathcal{O}(z^{-2}),\\
&& \lim_{z \rightarrow \infty} R(z) = \frac{N_0 \ell}{2}\Big(z+\frac{1}{z}\Big) + \mathcal{O}(z^{-2})\\
&& \lim_{z \rightarrow \infty} P_\Lambda(z) =L_0 z^2  +L_1 z +L_2+ \mathcal{O}(z^{-1}),\\
&& \lim_{z \rightarrow \infty} P_R(z) =R_0 z^2  +R_1 z+R_2+ \mathcal{O}(z^{-1})\,,
\ea 
\end{subequations}
for some to-be-determined constants $L_i$ and $R_i$.  
The constant $\ell$ was defined in the previous section, and $N_0=N_0(\beta,\gamma,Gm,\ell)$ is the asymptotic value of the lapse function \eqref{N}. In fact, assuming the above estimates, $N$ assumes the following asymptotic form:
\be \label{asn}
\lim_{z\rightarrow \infty}N(z) = N_0 + \frac{N_1}{z} + \frac{N_2}{z^2} + \mathcal{O}(z^{-3}),
\ee
where it is straightforward to show that $N_1= L_1 \tilde{N}_1(\beta,\gamma,Gm,\ell)$ and $N_2= L_1 ^2 \tilde{N}_2(\beta,\gamma,Gm,\ell) + (L_2-2L_0)\hat{N}_2(\beta,\gamma,Gm,\ell)$. 
For later convenience, we choose to parameterize 
\be
\ell:= \frac{2 G m}{\xi}\,,
\ee
for some positive dimensionless  constant $\xi$ which will be determined below.

With Eqs. \eqref{ansatz} and \eqref{asn} in hand, the  metric in the asymptotic limit becomes
\ba \la{dSmetric}
\lim_{z \rightarrow  \infty} g_{ab}dx^a dx^b
&=& - \frac{\ell^2}{z^2}\Big[N_0 ^2 + \frac{2 N_0 N_1}{z} + \frac{N_1 ^2 + 2 N_0 N_2}{z^2} + \mathcal{O}(z^{-3})\Big]dz^2 + \frac{1}{4}\Big[z^2 - 2+ \mathcal{O}(z^{-1})\Big]dx^2 \n\\
&& \ + \frac{N_0 ^2 \ell^2}{4}\Big[z^2 + 2 + \mathcal{O}(z^{-1})\Big]d \Omega^2.
\ea 
We will show in the subsequent section that with $N_1$ and $N_2$ vanishing, the above metric satisfies the criterion for the asymptotically Schwarzschild--de Sitter metrics as stipulated in \cite{Ashtekar:2014zfa} with a cosmological constant term given by (see Appendix  \ref{App:deSitter} for more details)
\be\la{CC}
\lambda=\frac{3}{N_0^2\ell^2 }\,.
\ee  
Fortunately, the vanishing of $N_1$ is a direct consequence of the requirement that $L_1=R_1=0$ which we demonstrate below. To arrange for the vanishing of $N_2$, it turns out that we must additionally impose $L_2=2 L_0$ as requiring $\hat{N}_2=0$ would be in conflict with Eqs. \eqref{xi1} and \eqref{xi2} below.
We shall use this latter relationship between $L_0$ and $L_2$ to simplify the ensuing equations.

Let us turn our attention to solving the scalar constraint and the three selected dynamical equations. Note that we have a total of four equations, which must be solved in vicinity of $z=\infty$ for up to three orders in $z$, in consistency with the orders kept in Eq. \eqref{ansatz}. That leaves us with a total of twelve algebraic equations between eleven a priori free parameters: $L_0$ (or $L_2$), $L_1$, $R_0$, $R_1$, $R_2$, $\gamma$, $\xi$, $\nu$
$\delta$, $\delta_x$, and $\beta$. At first glance,  this system of equations appears to be over-determined. We will see, however, that one of these equations is already zero and two other vanish if we require $L_1=0$. 

To begin, let us insert $R$ and $P_\Lambda$ in \eqref{ansatz} into Eq. \eqref{PLdot-CS} and find the following order-by-order algebraic equations:
\ba \label{PLorder}
&&{\rm order}\ z: L_0 - \frac{G m^2 N_0 ^2 [\pi - \arccos{(\xi)}]}{\beta \gamma \xi^2}=0,\n\\
&&{\rm order}\ z^0: 0, \n\\
&&{\rm order}\ z^{-1}: \pi h_0\Big[\frac{\sqrt{2 \pi} \beta  \xi^2 L_0}{8 G m^2 N_0 ^2}\Big]\bigg\{-8 \ell_p ^2 \beta^2 \gamma^2 \xi^2 \delta+ \delta_x \Big[2 \ell_p ^2 \gamma^2 \big(-4[\nu-3]\beta^2 + \ell_p ^2 \delta \big)\xi^2 + 2 G^2 m^2 N_0 ^2 \beta^2 \delta \Big] \bigg\} \n\\
&& \hspace{1.8 cm}+ 2 \sin{\Big[\frac{\sqrt{2 \pi }\beta \xi^2 L_0}{8 G m^2 N_0 ^2}\Big]}\bigg\{-8 \ell_p ^2 \beta^2 \gamma^2 \xi^2 \delta+ \delta_x \Big[\ell_p ^2 \gamma^2 \big(8[3\nu-1]\beta^2 + \ell_p ^2 \delta \big)\xi^2 + 2 G^2 m^2 N_0 ^2 \beta^2 \delta \Big]\bigg\}=0.
\ea
The order $z$ equation can be solved for $L_0$. We will see in Appendix \ref{sec:AppB} that the vanishing of the order $z^0$ equation is due to the fact that there is no $z^0$ term in the expansion for $R$. The last equation, on the other hand, dictates a relationship between $\xi$ and the five quantum parameters $\beta$, $\gamma$, $\delta$, $\delta_x$, and $\nu$. Note that this equation was already simplified using $L_2=2L_0$.

With $P_\Lambda$ determined to the desired order, we move on to solving for the constants $R_i$ in $P_R$ using Eq. \eqref{Rdot-CS}, which results in the following algebraic equations:
\ba \label{Rorder} 
&&{\rm order}\ z: \xi + \cos{\Big[\frac{\beta \gamma \xi}{G m^2 N_0 ^2} (2 G m R_0 N_0 - L_0 \xi)\Big]}=0,\n\\
&&{\rm order}\ z^0: \frac{\beta \gamma}{m N_0} \sin{\Big[\frac{\beta \gamma \xi}{G m^2 N_0 ^2} (2 G m R_0 N_0 - L_0 \xi)\Big]} (2 G m N_0 R_1 - L_1 \xi)=0, \n\\
&&{\rm order}\ z^{-1}: -2 G m N_0  - 2 \beta \gamma G  R_2  \sin{\Big[\frac{\beta \gamma \xi}{G m^2 N_0 ^2} (2 G m R_0 N_0 - L_0 \xi)\Big]}  - \frac{2 \ell_p ^2 \gamma^2 \xi^2}{G m N_0 \beta^2 \delta \delta_x \Big(2 \sin{\Big[\frac{\sqrt{2 \pi }\beta \xi^2 L_0}{8 G m^2 N_0 ^2}\Big]} + \pi h_0 \Big[\frac{\sqrt{2 \pi }\beta \xi^2 L_0}{8 G m^2 N_0 ^2}\Big]\Big)} \n\\
&&\hspace{1.8 cm} \times \Big\{\pi h_0\Big[\frac{\sqrt{2 \pi }\beta \xi^2 L_0}{8 G m^2 N_0 ^2}\Big] \Big(-4 \beta^2 \delta + \delta_x [\ell_p ^2 \delta-4(\nu-3) \beta^2]\Big) +\sin{\Big[\frac{\sqrt{2 \pi }\beta \xi^2 L_0}{8 G m^2 N_0 ^2}\Big]}\Big(-8 \beta ^2 \delta + \delta_x [8 \beta^2 (3 \nu -1) + \ell_p ^2 \delta]\Big)\Big\}\n\\
&& \hspace{1.8 cm}=0,
\ea
where the last equation was simplified using the previous two equations.
As in \eqref{PLorder}, we solve these algebraic equations for $R_0$, $R_1$, and $R_2$. Note that $R_1$ is entirely dependent on $L_1$ \footnote{Requiring $\sin{\big[\frac{\beta \gamma \xi}{G m^2 N_0 ^2}(2Gm R_0 N_0 - L_0 \xi)\big]}=0$ in lieu of $2G m N_0 R_1 - L_1 \xi = 0$ leads to inconsistencies in the subsequent algebraic equations.}. Also, it is straightforward to confirm that $R_2=0$ by virtue of the order $z^{-1}$ term in Eq. \eqref{PLorder}. 

At this stage, we have a complete set of asymptotic solutions for all phase space variables in terms of $Gm$, $\xi$, $\ell_{\rm \va p}$, and the five quantum parameters $\beta$, $\gamma$, $\delta$, $\delta_x$, and $\nu$. The next step is to ensure the consistency of the scalar constraint equation as well as Eq. \eqref{Ldot-CS} to the desired order in $z$. As we will see shortly, this consistency mandates fine tuning for most of our quantum parameters. 
Let us consider the order $z^0$ term in $\mathcal{H}_{\rm int} ^{\rm \va IV+CS}/R^2 \Lambda=0$:
\ba \label{xi1}
&&\pm \frac{\sqrt{2 \pi}}{8 \gamma} \sqrt{1-\xi^2} \big(2 \sin{[\iota]}+ \pi h_0[\iota]\big) + \pi \sin{[\iota]} h_0[\iota ]=0,
\ea
where $\iota := \sqrt{2 \pi} [\pi-\arccos{(\xi)}]/(8 \gamma)$ and we used the order $z$ equations in \eqref{PLorder} and \eqref{Rorder}\footnote{Note that Eq. \eqref{Rorder} does not fix the sign of $\sin{\Big[\frac{\beta \gamma \xi}{G m^2 N_0 ^2} (2 G m R_0 N_0 - L_0 \xi)\Big]}$ in Eq. \eqref{xi1}, requiring us to account for both signs at this stage.}. This equation couples $\xi$ to  $\gamma$. Another such equation is the order $z^0$ term in the expansion of Eq. \eqref{Ldot-CS}:
\ba \label{xi2}
&&  -2 \xi - \frac{\pi \big(2 \sin^2{[\iota]} h_{-1}[\iota\big] + \pi \cos{[\iota]} h_0 ^2[\iota]\big)}{\big(2 \sin{[\iota]}+ \pi h_0[\iota] \big)^2}=0.
\ea
These two algebraic equations admit two sets of solutions for $\xi$ and $\gamma$ which we list below. 

Moving on to the order $z^{-1}$ term in $\mathcal{H}_{\rm int} ^{\rm \va IV+CS}/R^2 \Lambda=0$, we find the following equation after imposing $2Gm N_0 R_1 - L_1 \xi =0$:
\ba 
&& \xi L_1 \Big\{\pm \frac{\sqrt{2 \pi}}{8 \gamma} \sqrt{1-\xi^2} \big(2\cos{[\iota]}+\pi h_{-1}[\iota]\big) + \pi \big(\sin{[\iota]} h_{-1}[\iota] + \cos{[\iota]} h_0 [\iota]\big) \Big\}=0,
\ea
where $h_{-1}$ is the Struve function of order $-1$. The expression inside the above parenthesis is not implied by Eqs. \eqref{xi1} and \eqref{xi2}, which necessitates the vanishing of $L_1$. In addition, it follows with no difficulty that the vanishing of $L_1$ results in the vanishing of $R_1$ as well as the order $z^{-1}$ term in Eq. \eqref{Ldot-CS}. 

Finally, let us examine the order $z^{-2}$ term in $\mathcal{H}^{\rm \va IV+CS}_{\rm int} /R^2 \Lambda = 0$ and Eq. \eqref{Ldot-CS}. Starting with the former, setting $R_1=L_1=0$, $L_2=2 L_0$, and making use of the order $z$ equations in \eqref{PLorder} and \eqref{Rorder}, we find
\ba \label{Hz2}
&& 2 \pi \beta^4 \delta \delta_x \pm 8 \sqrt{2 \pi (1-\xi^2)} \ell_p ^2 \gamma \sin{[\iota]} \big(-8 \beta^2 \delta + \delta_x [8(-1+3 \nu) \beta^2 + \ell_p ^2 \delta]\big) + 8 \gamma \ell_p ^2 h_0 [\iota] \Big(\pm \sqrt{2 (1-\xi^2)} \pi^{3/2}  \big(-4 \beta^2 \delta \n\\
&&+\delta_x [-4(-3+\nu)\beta^2 + \ell_p ^2 \delta] \big) + 32 \gamma \sin{[\iota]} \big(3 \pi \beta^2 \delta - \delta_x [\pi (1+\nu) \beta^2 - 8 \ell_p ^2 \gamma^2 \delta]\big) \Big)=0.
\ea
As for Eq. \eqref{Ldot-CS}, we end up with the following algebraic equation after incorporating the previously mentioned simplifications together with the last equation in \eqref{PLorder}:
\ba \label{Lz2} 
&& \pm 4 \sqrt{2 (1-\xi^2)} \pi^{3/2} \ell_p ^2 \gamma  \delta_x [32 (-1+\nu) \beta^2 - \ell_p ^2 \delta]  \big(\sin{[\iota]} h_{-1}[\iota] - \cos{[\iota]} h_0 [\iota]\big) +  h_{-1}[\iota]  \Big(\pi^2 \beta^4 \delta \delta_x - 256 \ell_p ^2 \gamma^2 \sin^2{[\iota]} \n\\
&&\times \big[3 \pi \beta^2 \delta - \delta_x (\pi (1+\nu) \beta^2 - 8 \ell_p ^2 \gamma^2 \delta)\big]\Big) + 2 \pi  \cos{[\iota]} \Big(\beta^4 \delta \delta_x - 64 \ell_p ^2 \gamma^2 h_0 ^2[\iota]  \big(3 \pi \beta^2 \delta - \delta_x[\pi (1+\nu)\beta^2 - 8 \ell_p ^2 \gamma^2 \delta]\big)\Big) =0.\n\\
\ea
We solve the above equation for $\delta$, the order $z^{-1}$ equation in \eqref{PLorder} for $\delta_x$, and Eq. \eqref{Hz2} for $\nu$. 
The results found for $\xi$ and $\gamma$ and the coherent state parameters $\delta$, $\delta_x$, and $\nu$ are reported in Table \ref{tabparam} below. 
\begin{table}[h]
\centering
\begin{tabular}{cc|c|c|}
\cline{3-4}
& &  $\sin{[\pi-\arccos{(\xi)}]<0}$& $\sin{[\pi-\arccos{(\xi)}]>0}$  \\ \cline{1-4}
\multicolumn{1}{|}{}&$\xi$&  $ 0.957$ &  $ 0.974$ \\ \cline{1-4}
\multicolumn{1}{|}{}&$\gamma$&  $0.227$ & $ 0.274$ \\ \cline{1-4}
\multicolumn{1}{|}{}&$\nu$&  $ 1.802$ & $ 1.802$ \\ \cline{1-4}
\multicolumn{1}{|}{}&$\delta$& 
${2.916}/{\beta^2}
  + \mathcal{O}(\beta^{-6})$ 
  & 
   ${1.458}/{\beta^2}
 + \mathcal{O}(\beta^{-6})$ 
  \\ \cline{1-4}
\multicolumn{1}{|}{}&
$\delta_x$&
 ${0.729}/{\beta^2} 
 + \mathcal{O}(\beta^{-6})$ 
 & 
  ${0.729}/{\beta^2} 
+ \mathcal{O}(\beta^{-6})$
 \\ \cline{1-4}
\end{tabular}
\caption{These are the values found for the parameters of the model that were tuned to bring about an asymptotically de Sitter geometry for the black hole interior region. We have considered both signs for $\sin{\Big[\frac{\beta \gamma \xi}{G m^2 N_0 ^2} (2 G m R_0 N_0 - L_0 \xi)\Big]}$ whose argument was further simplified to $\pi- \arccos{(\xi)}$ using the order $z$ equations in \eqref{PLorder} and \eqref{Rorder}. Also, in the expansions for $\delta$ and $\delta_x$ we are mainly after the $\beta \gg 1$ limit.}
\label{tabparam}
\end{table}

As a consistency check, we must ensure that the values found above for the coherent state parameters 
satisfy the condition given in Eq. \eqref{deltaj}. Inserting the expressions for $\delta$ and $\delta_x$ that are reported in Table \ref{tabparam}, it is immediate to see that
\be
\delta_r \tilde j_r^2\approx \delta_\theta\tilde j_\theta^2 \approx \delta_\varphi \tilde j_\varphi^2 \approx 
 \frac{R^2}{ \beta^2  } = \epsilon^{-2}\,,
\ee
which attests to the validity of our series expansion for the coherent state corrections.

To summarize, we have shown that the effective dynamics generated by the Hamiltonian \eqref{heff-IV+CS} admits a solution in the interior region that in the limit $z \rightarrow \infty$ assumes the following form
 \ba \label{esssol}
&&N = -\frac{\sqrt{2 \pi} \beta}{G m \big(8\sin{[\iota]}+ 4 \pi h_0[\iota]\big)} + \mathcal{O}(z^{-3}), \n\\ 
&& \Lambda (z) = \frac{1}{2} \Big(z- \frac{1}{z}\Big) + \mathcal{O}(z^{-2}),\n\\
&& R(z) = \frac{\sqrt{2 \pi} \beta}{\xi \big(8 \sin{[\iota]} + 4 \pi h_0 [\iota]\big)} \Big(z+\frac{1}{z}\Big) + \mathcal{O}(z^{-2}), \n\\
&&P_\Lambda(z) = \frac{\pi \beta (\pi-\arccos{[\xi]})}{8 G \gamma \xi^2 \big(2 \sin{[\iota]} + \pi h_0 [\iota] \big)^2} (z^2 + 2) + \mathcal{O}(z^{-1})\n\\
&&P_R (z) = - \frac{\sqrt{\frac{\pi}{2}}(\pi - \arccos{[\xi]})}{2 G \gamma \xi \big(2 \sin{[\iota] + \pi h_0 [\iota]}\big)} z^2 + \mathcal{O}(z^{-1}),
\ea
with parameters $\xi$ and $\gamma$ (and $\iota$) taking any of the values in Table \ref{tabparam}. We will show in Sec. \ref{gpscri} that a metric with the above listed components satisfies the criterion for an asymptotically Schwarzschild--de Sitter metric with a  cosmological constant \eqref{CC} that is given by 
\ba \la{lambda}
\lambda &=& \frac{6 \xi^2 \big(2 \sin{[\iota]} + \pi h_0 [\iota]\big)^2}{\pi \beta^2} \n\\
&=& \frac{3}{256 \gamma^2 \ell_{\va P} ^2 j} \bigg(\frac{2 \sin^2{[\iota]} h_{-1}[\iota] + \pi \cos{[\iota]} h_0 ^2 [\iota]}{2 \sin{[\iota]} + \pi h_0[\iota]}\bigg)^2,
\ea
where we replaced $\beta$ and $\xi$ using Eqs. \eqref{epsilons} and \eqref{xi2} respectively.
It should be noted  that $\lambda$ is inversely proportional to the quantum spin number $j$ only and it is thus purely of quantum gravitational origin.

A few remarks are in order here. First, the most striking feature of the solution set \eqref{esssol} is the required value of $\gamma$. In fact, our analysis has shown that an asymptotically de Sitter universe extending the black hole interior spacetime beyond the classical singularity is predicted by LQG for the  Barbero--Immirzi parameter  corresponding to either one of the following two numerical values (the superscript ${\rm dS}$ stands for de Sitter):
\be\la{gamma-dS}
\gamma^{\rm {\va dS}}_1\approx0.227 \,,\quad \gamma^{\rm {\va dS}}_2\approx 0.274\,.
\ee
In the context of black hole physics, this is not the first time that two different numerical values for $\gamma$ have been derived. Indeed, it is well known that the black hole entropy calculations in LQG require fine tuning the Barbero--Immirzi parameter in order to recover the numerical factor $1/4$ in the Bekesntein--Hawking entropy-area formula \cite{Bekenstein:1973ur, Hawking:1974sw}. The LQG calculation relies on the notion of {\it isolated horizons} \cite{Ashtekar:2004cn}. The original derivation was done in the $\U(1)$ framework and developed in  \cite{Ashtekar:1999wa, Ashtekar:2000eq, Meissner:2004ju, Domagala:2004jt}, where a partial gauge fixing of the horizon geometrical structure was performed. Later on, this symmetry reduction was replaced with a full $\SU(2)$-invariant analysis of spherically symmetric isolated horizons, with the resultant entropy calculations reported in \cite{  Engle:2010kt, Perez:2010pq, Engle:2011vf, Agullo:2009eq}. Previous insights from  \cite{Kaul:1998xv, Ghosh:2006ph} were incorporated in the latter set of papers (see \cite{DiazPolo:2011np} for a review of both frameworks). 
The two formulations lead to different numerical values for the Barbero--Immirzi parameter.
These two values were precisely derived in \cite{Agullo:2010zz} where the authors performed a detailed analysis relying on number theoretic and combinatorial methods. These values were found to be 
 \be\la{gamma-E}
\gamma^{\rm {\va E}}_1\approx0.237 \,,\quad \gamma^{\rm {\va E}}_2\approx 0.274,\,
\ee
for $\U(1)$ and $\SU(2)$ respectively (the superscript ${\rm E}$ denotes entropy).
We see that the value $\gamma^{\rm {\va dS}}_1$  corresponding to the   case $\sin{[\pi-\arccos{(\xi)}]<0}$ listed in Table \ref{tabparam} is surprisingly close to the value $\gamma^{\rm {\va E}}_1$ obtained from the $\U(1)$ black hole entropy counting in LQG. Even more remarkable is the exact matching of $\gamma^{\rm {\va dS}}_2$ for the $\sin{[\pi-\arccos{(\xi)}]>0}$ case and the $\SU(2)$ counting value  $\gamma^{\rm {\va E}}_2$.

  While our construction of a quantum black hole geometry in \cite{Alesci:2019pbs} is compatible with the quantum isolated horizon framework  and the standard LQG entropy counting (a priori both in the $\U(1)$ and the $\SU(2)$ formulations), none of those ingredients have been used in our derivation. What we have presented here is a genuinely independent derivation of two numerical values for $\gamma$ that are surprisingly close to what was previously found in a very different context. Most importantly, however, is the fact that this derivation emerged from searching for metrics with distinctive asymptotic characters that would not have existed in this context if it were not for the strong quantum gravitational effects in which the Barbero--Immirzi parameter is known to play a pivotal role. In fact, the reason as to why fixing $\gamma$ in the black hole entropy calculation has been a  controversial topic in the previous literature (see, e.g., \cite{Jacobson:2007uj, Ghosh:2011fc, Frodden:2012dq, Ghosh:2013iwa, Pranzetti:2013lma, Bodendorfer:2013hla, Ghosh:2014rra, Pranzetti:2014tla, Achour:2014eqa, BenAchour:2016mnn, Oriti:2018qty}) is due to the fact that the introduction of $\gamma$ has no implications for the bulk classical dynamics and thus is not expected to play a role in recovering the Bekenstein--Hawking entropy formula \footnote{However, the Barbero--Immirzi parameter plays an important role in the analysis of the boundary symmetry algebra of gravity at the classical level, as revealed by the edge modes formalism \cite{Freidel:2020xyx, Freidel:2020svx, Freidel:2020ayo}, and this can indeed indicate direct implications for black hole physics in the semi-classical regime already.}; the  numerical coefficient there is derived from the black hole Hawking temperature \cite{Hawking:1974sw}, which is a result of QFT on a classical curved background. In other words, no quantum gravity ingredient is required to derive the factor $1/4$ in the black hole entropy-area relation. Thus, it remains mysterious as to why the Barbero--Immirzi parameter needs to be fixed in the first place. On the other hand, the emergence of an asymptotically de Sitter universe replacing the Schwarzschild black hole singularity is a result that has no semi-classical analog and it is purely quantum gravitational in origin. In this sense, our result represents a striking coincidence and, at the same time, a long sought-after confirmation of the special physical meaning for the numbers \eqref{gamma-dS}.
 We should mention here, however, that our numerical investigations indicate that we always end up with the positive sine function in Table \ref{tabparam}  should we integrate the dynamical system starting with an initial data set that is close to that of a Schwarzschild black hole in vicinity of its event horizon. Hence, while $\gamma_1 ^{ds}$ is a possibility not excluded by the dynamics, it is $\gamma_2 ^{ds}$ that is relevant for black hole physics.

Second, we want to stress that including the coherent state corrections in \eqref{corrections} is necessary for recovering all of the terms for the metric variables given in Eq.  \eqref{ansatz}. Terms of order $1/z$ in $R$ and $\Lambda$ are of crucial importance for otherwise the corresponding metric would fail to meet all the criteria introduced in \cite{Ashtekar:2014zfa} for an  asymptotically de Sitter spacetime. In particular, while  the order $z$ terms in $R$ and $\Lambda$ are enough for the curvature invariants to resemble those of a de Sitter metric, the subleading terms play a key role in satisfying the proper fall-off conditions for the effective stress-energy tensor  $T^{\rm eff}_{ab}$
 defined by the solution \eqref{dSmetric} with $\lambda$ given in \eqref{CC} (see next subsection for more details on the asymptotic geometry).  This shows how, in this specific case studied here, the geometrical properties of an asymptotically de Sitter metric fixes all the ambiguities left in the definition of the coherent states \eqref{co-st}, with the only free parameter left being the spin number $j$ entering  the definition of $\beta$ in \eqref{epsilons}. 


\subsection{Geometry at $\mathcal{I}^+ _{\rm int}$} \label{gpscri}

Having found the metric \eqref{dSmetric} as a solution to the effective Hamiltonian system, it is now  simple to show that the black hole interior satisfies the criterion for asymptotically Schwarzschild--de Sitter spacetimes. To begin, let us define the conformal factor $\omega := 1/z$ and re-scale the metric \eqref{dSmetric} with $N_1=N_2=0$ by $\omega^2$ to find \footnote{Though not shown in this paper, it is straightforward to see that powers of $z$ in the asymptotic expansions for the phase space variables are separated by $2$. Therefore, the corrections in square parenthesis are in fact $\mathcal{O}(\omega^4)$. This fact has no bearing on our discussion in this section.}
\ba \la{confmet}
\tilde{g}_{ab} dx^a dx^b := \omega^2 g_{ab} dx^a dx^b &=& - N_0 ^2 \ell^2 \Big[1+ \mathcal{O}(\omega^3)\Big] d \omega^2  + \frac{1}{4}\Big[1-2 \omega^2 + \mathcal{O}(\omega^3)\Big] dx^2\n\\
&\ \ &+\frac{N_0 ^2 \ell^2}{4}\Big[1+2 \omega^2 + \mathcal{O}(\omega^3)\Big] d \Omega^2,
\ea
where $N_0$ is the leading order term for the lapse function and is given in Eq. \eqref{esssol} and $\ell = 2 G m /\xi$ with the value of $\xi$ listed in Table \ref{tabparam}. It is clear that the conformally rescaled metric $\tilde{g}_{ab}$ is smooth everywhere, including at the conformal boundary given by $\omega=0$. We call this boundary the interior scri plus and denote it by $\mathcal{I}^+ _{\rm int}$. Note that $d \omega$ is no-where vanishing along $\mathcal{I}^+ _{\rm int}$. The intrinsic metric on this space-like hypersurface is that of a Euclidean cylinder,
\be 
\tilde{h}_{ab} dx^a dx^b  = \frac{1}{4} dx^2 + \frac{N_0 ^2 \ell^2}{4} d \Omega^2.
\ee
Evidently, $\mathcal{I}^+ _{\rm int}$ is geodesically complete with respect to $\tilde{h}_{ab}$.

\begin{figure}[h]
\centering
\includegraphics[width=0.6\textwidth]{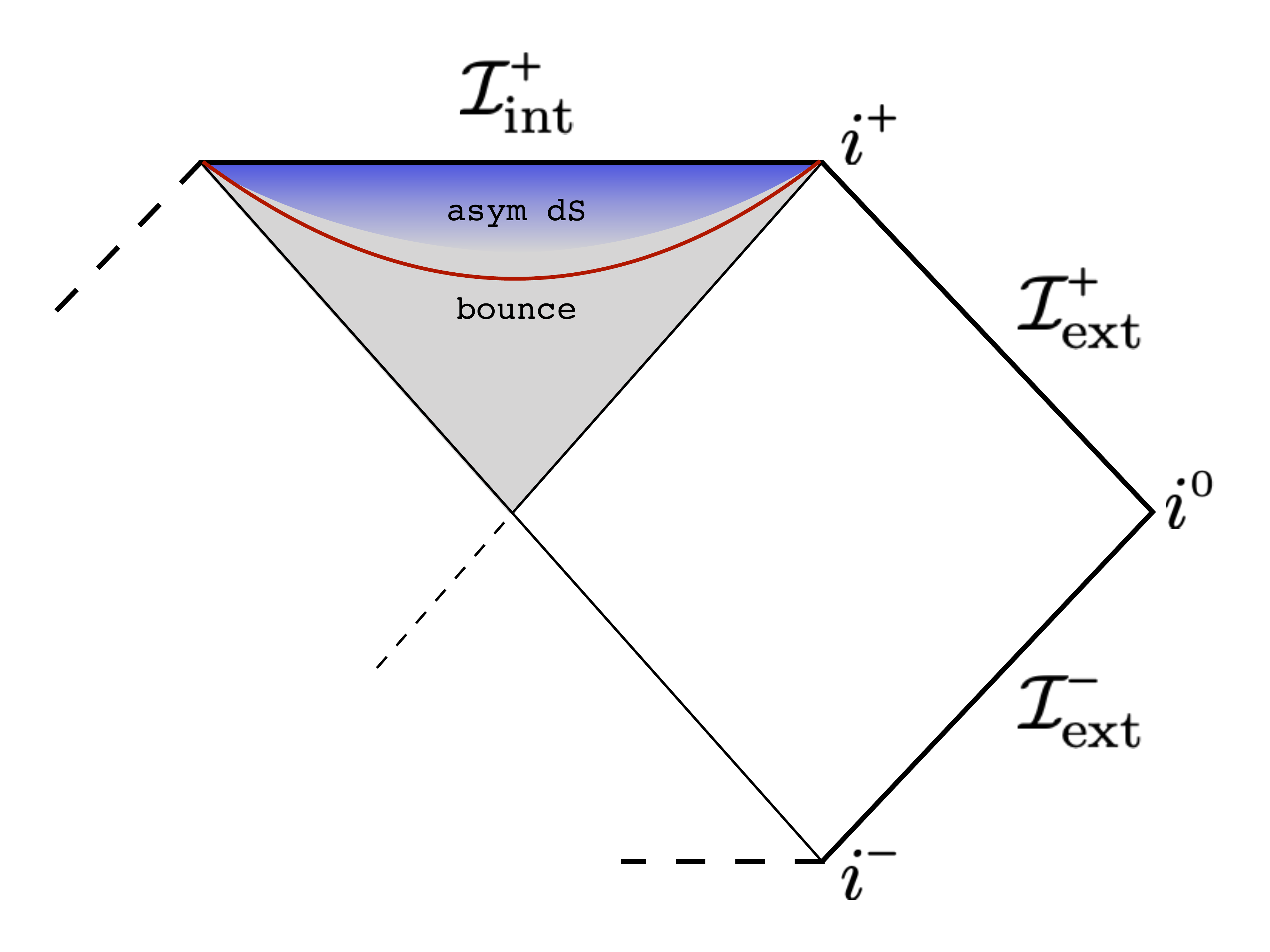}
\caption{This is our proposed Penrose diagram for the Schwarzschild spacetime based on our interior quantum gravity model. The grey fading into blue shaded region is the interior of the black hole; it is causally bounded by the future event horizon to its past and the space-like conformal boundary $\mathcal{I}_{\rm int} ^+$ to its future. The Schwarzschild singularity is replaced by a transition surface, denoted by the red curve, in proximity of the region where the classical curvature becomes Planckian.  The resultant physical metric $g_{ab}$ is smooth everywhere. Asymptotically, the effective interior spacetime metric approaches the de Sitter metric \eqref{dS3}.}
\label{penrose}
\end{figure}

It remains to see how the components of the effective stress-energy tensor, 
\be 
T^{\rm eff}_{ab} :=  \frac{1}{8 \pi G} \big[G_{ab} + \lambda g_{ab}\big]\,,
\ee
behave as $\omega \rightarrow 0$. Here $G_{ab} = R_{ab} - (R/2) g_{ab}$ is the Einstein tensor and $\lambda$ is the emergent cosmological constant which is  explicitly given in Eq. \eqref{lambda}. A quick calculation reveals 
\ba \la{Teff}
&& \lim_{\omega \rightarrow 0 }T^{\rm eff}_{\omega \omega} = \mathcal{O}(\omega), \ \ \ \ \lim_{\omega \rightarrow 0 }T^{\rm eff}_{xx} = \mathcal{O}(\omega^2), \ \ \ \ \lim_{\omega \rightarrow 0}T^{\rm eff}_{\theta \theta} = \frac{1}{\sin^2{(\theta)}} \lim_{\omega \rightarrow 0} T^{\rm eff}_{\phi \phi} = \mathcal{O}(\omega^2).
\ea 
Since the components of the effective stress-energy tensor vanish at least as fast as $\omega$ as $\omega \rightarrow 0$, and due to $\mathcal{I}^+ _{\rm int}$ being homeomorphic to $\mathbb{R}\times \mathbb{S}^2$, the interior spacetime is asymptotically Schwarzschild--de Sitter per the criteria stipulated in \cite{Ashtekar:2014zfa}. The Penrose diagram for the Schwarzschild spacetime is now replaced by the one given in Fig. \ref{penrose}.

We end this subsection with a few remarks. First, it is an intriguing fact that the energy stored in the gravitational field as perceived by external observers is rarefied by the quantum effects in the interior region. More precisely, even though a non-zero Bondi mass can be computed at any $\mathbb{S}^2$ cross section of $\mathcal{I}_{\rm ext}$, there is no non-zero gravitational mass in the interior region. In fact, gravitational charges for asymptotically de Sitter spacetimes with conformally flat intrinsic metrics at scri can be unambiguously defined using \footnote{See \cite{Ashtekar:2014zfa} for the derivation and discussions.}
\be 
\oint_{\mathcal{C}} \mathcal{E}_{ab} \mathcal{K}^a \hat{x}^b d^2 V
\ee
for any $\mathbb{S}^2$ cross section $\mathcal{C}$ of scri. Here $\mathcal{K}^a$ is a Killing vector field of $g_{ab}$, $\hat{x}^a$ is the unit normal to $\mathcal{C}$, and $\mathcal{E}_{ab}$ is the electric part of the rescaled Weyl tensor that is defined as $ C_{acbd}n^c n^d /\omega$ with $n^a$ being the unit normal vector to the constant $\omega$ surfaces. A quick calculation shows that $\mathcal{E}_{ab}$ vanishes at $\mathcal{I}^+ _{\rm int}$. 

Second, as the interior region becomes nearly isometric to the de Sitter space in vicinity of $\mathcal{I}^+ _{\rm int}$, one might expect that local asymptotic observers see a high degree of homogeneity and isotropy in their observable universe. However, as in the case of the de Sitter space in static coordinate patch (see Figure \ref{dS-static}), the global topology of $\mathcal{I}^+ _{\rm int}$ 
prevents the existence of all finite symmetry transformations except those generated by the four Killing vector fields that we had started with (even in an approximate sense). In other words, the symmetry Lie group of the interior region remains isomorphic to $\mathbb{R}  \rtimes {\rm SO}(3)$, just as that of the exterior region. 
%

\section{Emergent cosmological constant}\la{sec:CC}

The emergence of a non-vanishing cosmological constant in the asymptotic post-bounce region from quantum gravitational effects is the most striking feature of this model. What is even more fascinating is that this is achieved by selecting a specific value for the Barbero--Immirzi parameter which exactly coincides with the one derived from the $\SU(2)$ black hole entropy calculation \footnote{As pointed out at the end of Section \ref{sec:series}, only $\gamma^{\rm {\va dS}}_2$ is relevant to the Schwarzschild black hole.}. The emergent cosmological constant \eqref{lambda} is a function of the average spin number $j$ associated to the links of the cuboidal graph that we introduced in order to construct the quantum reduced Hilbert space \cite{Alesci:2018loi}. After inserting the numerical values for $\gamma$ and $\xi$ listed in the second column of  Table. \ref{tabparam}, it becomes
\be\la{lambda-j}
\lambda=\frac{0.06}{  \ell_{\va P}^2 j }\,.
\ee

At first sight, $\lambda$ appears to be unconstrained 
since $j$ is a priori a free parameter that enters the effective solution \eqref{esssol}. Regarding $j$, all that we have demanded in our analysis so far is the restriction that $1\ll\sqrt{j} \ll G m/ \ell_{\va P} $ on $\Sigma$ where we set the initial conditions, which can be taken to be arbitrarily close to the event horizon. Meanwhile, the lower bound condition $j\gg1$ comes from the requirement  \eqref{deltaj} to guarantee sufficient peakedness for the coherent states \eqref{co-st}. We do not need to require $j$ to be large. In fact, the coherent states that we used to derive the effective Hamiltonian \eqref{heff-IV+CS} are  sufficiently peaked  for $j$ of order $\sim 100$ (see, e.g., \cite{Livine:2007vk}). 
On the other hand, the upper bound condition $\sqrt{j} \ll G m/ \ell_{\va P}$ comes from the requirement $\epsilon\ll1$ on $\Sigma$ near the event horizon, where quantum gravity effects are expected to be negligible. It can be checked that small variations of the Schwarzschild initial data satisfy the scalar constraint equation ${\mathcal H}^{\rm \va IV+CS}_{\rm int} =0$ with small error for spins as high as $j\sim G m/ \ell_{\va P}$. We warn the reader, however, that  instead of considering the {\it average} spin in Eqs. \eqref{Ne}-\eqref{ep2}, one could consider a {\it coarse grained spin} $j^{cg}$ for a number of cells $\mathcal N^{cg}< \mathcal N$  without any change in our equations, except that in this case $j^{cg}$ can be this large.  In fact, while individually and on each link small values of the spins $j^p$ may be more likely, the coarse grained value $j^{cg}$  can be large for large horizon area. In fact, an expectation value for $j$  (in the rest of the paper we will just refer to a {\it collective} $j$ that can be either the averaged or the coarse grained one) scaling with the black hole mass was derived in \cite{Ghosh:2013iwa} where the authors considered  fermionic statistics for a gas of punctures 
within the quantum isolated horizons framework.
Moreover, it has been argued in \cite{Sahlmann:2001nv} that such mesoscopic scale  should correspond to the regime where   the continuum and classical limits coincide.
Therefore, for astrophysical black holes, it would seem  that there is a wide range of collective spin numbers that one can use without risking the emergence of large quantum effects in vicinity of the event horizon.

That being said, a more stringent upper bound on $j$ can be set by demanding that quantum effects remain subdominant up until the moment when the spacetime curvature becomes Planckian. This can be quantified by requiring that the bounce occurs where the
Kretschmann scalar  \eqref{K} of the Schwarzschild metric becomes Planckian, namely by demanding 
\be
\mathcal{K}_c(R_b) \sim \frac{1}{\ell_{\va P}^4}\,,
\ee
where $R_b$ is the minimum value reached by the metric function $R$ at the instant of bounce. Our numerical analysis points to the following dependence of $R_b$ on the classical black hole mass and the collective spin number:
\be
R_b(m,j)\sim  (Gm)^\frac{1}{3} (\ell_{\va P}^2j)^\frac{1}{3}\,.
\ee
If one wants to strictly confine quantum effects to within the Planckian curvature region, then this relation could be used to argue for  an upper bound on $j$ of order $\sim 10^6$ or so for large enough black holes. These considerations should serve as a reminder to the reader that the effective theory framework leaves $j$, and thereby $\lambda$, largely unconstrained. 

What complicates the matter even further is the interpretation of the effective continuum dynamics from the point of view of the fundamentally discrete structure of the quantum theory. In fact,  tensions appear when trying to understand the pre-bounce contraction and the post-bounce expansion of the metric function $R$ from the point of view of the fixed graph structure used to derive the effective dynamics. In particular, the use of a non-graph-changing Hamiltonian in the definition of the microscopic dynamics would seem consistent with such contraction and expansion of space  only  if the quantum spin numbers associated to the links of the graph change at each step of the Hamiltonian constraint action. This strongly suggests that the collective spin $j$ should then undergo a renormalization flow, with its value possibly changing from what it is at $\Sigma$ where $z\rightarrow 1$ to an asymptotic one as $z\rightarrow\infty$. At the same time, this quantum number is an input from the full theory kinematical Hilbert space structure that is ``invisible'' to our semi-classical effective dynamics description. More precisely, the collective spin $j$ and  the number of plaquettes $\mathcal N$ (not necessarily the total one in the case of a coarse grained $j$) cannot be represented {\it separately} as classical phase space functions; it is only the combination $\ell^2_{\va P}  j \mathcal N^2=R^2$ that can be evolved by our effective dynamics. 

In other words,  from the point of view of the microscopic theory, the effective time evolution of the metric function $R(z)$ can be understood as the action of the Hamiltonian constraint changing, at each step, the quantum numbers of the spin network states. However, from the point of view of our effective description, we cannot discern whether it is $j$ that is changing, or $\mathcal N$, or any combination of the two: all we have access to is the  metric function $R$. This ambiguity can only be resolved by having a better control over the   microscopic  dynamics and the coarse-graining properties of the quantum physical states, for instance by following an approach similar to the one advocated in  \cite{Dittrich:2014ala} for the construction of physical states through iterative coarse graining methods.  Unfortunately, we are quite far from achieving this.


Nevertheless, in light of this discussion, one could contemplate a mechanism where, as a result of microscopic  quantum dynamics evolution, the initial value $j_i$ of the { collective} spin entering the construction of the coherent states (through the spatial geometry regularization structure as $\ell^2_{\va P}  j_i \, \mathcal N^2_i=R_{i}^2$) gets {\it renormalized}  by the initial number of cells $\mathcal N_i^2$, namely 
\be\la{jbar}
\bar j\sim j_i \, \mathcal N_i^2\sim G^2m^2/\ell_{\va P}^2\,, 
\ee
where we are using the initial condition   $ R_i \sim Gm$.
 In the resulting  extended spacetime, the asymptotic region after the bounce is then described by a near  de Sitter metric \eqref{esssol} with a {\it renormalized} positive cosmological constant (we restore the speed of light  $c$ here)
\be\la{CCbar}
\bar \lambda\sim \frac{c^4}{ G^2 m^2 }\,.
\ee

Let us now sketch a simple argument in favor of the rescaling proposed in Eq. \eqref{jbar} for the collective spin. The macroscopic universe is expected to behave classically once again in the post-bounce region and in vicinity of $\mathcal{I}^+ {\rm int}$. However, all curvature scalars in this territory eventually become proportional to $\lambda$ as $z \rightarrow \infty$ and thus
divergent in the classical limit $\hbar \rightarrow 0$. The only possibility to remove the explicit $\hbar$ dependence in the curvature scalars of the emergent quasi-de Sitter universe is to have it canceled by a non-analytic $\hbar$ dependence in the collective spin. However, as $\bar{j}$ is dimensionless, it ought to be given by the square of the ratio of some length scale over the Planck length. Given, that the only other length scale in this theory is $G m$,  we are left with no natural option but to rely on the rescaling behavior \eqref{jbar} in order to wind up with a classical macroscopic post-bounce universe.

If the expression \eqref{CCbar} for the renormalized cosmological constant can indeed be obtained in a renormalization-like process, then it is truly fascinating due to the following observation: Inserting the value for the observed mass ({\it i.e.} non-relativistic matter) in the universe in place of $m$ results in $\bar{\lambda}$ being on the same order of magnitude as the measured value of the cosmological constant of our universe, that is $ \lambda_{obs} \simeq 1.1 \times 10^{-52} \, {\rm m}^{-2}$. More precisely, since $m$ in \eqref{CCbar} has the interpretation of the black hole's gravitational mass as measured by a stationary observer near $\mathcal{I}_{\rm{ext}}$, if our universe hides behind a black hole event horizon, then this quantity as perceived by an observer in the mother universe should correspond to the mass of the matter content of our observable baby universe. If we insert in \eqref{CCbar} the value of the mass of baryonic matter \footnote{A contribution from dark matter to $\bar{\lambda}$ cannot be ruled out at this stage since we are neglecting constant factors of $\mathcal O (1)$ in the proposal \eqref{CCbar}, even though we limit our considerations to baryonic matter for the sake of the arguments presented here.
} as obtained from the cosmological parameters measured by the Planck Mission in 2018 \cite{Aghanim:2018eyx}, namely by setting $m\simeq 1.46 \times 10^{53}  \,{\rm kg} $, then we obtain $\bar \lambda\simeq 0.85 \times 10^{-52} \, {\rm m}^{-2} $!

Before leaving this intriguing observation as a starting point for future investigations about possible observables effects of our model, let us propose a more precise framework in which a relation like \eqref{CCbar}  could be obtained. To this end, the  basic observation is that even though a priori {\it all} values of $j_i$ and $ \mathcal N_i$ are allowed as long as  their product gives $R_{i}$, one would still need to introduce an initial {\it distribution}  $\rho(\mathcal N$) on the number of plaquettes. Is it possible to evolve this $\rho(\mathcal N)$ in absence of a graph changing Hamiltonian? The answer is indeed in the affirmative  if one considered the system $(\mathcal N ,j)$ as an open quantum system and treated the quantum degree of freedom $j$ as diffusing in a stochastic bath of $\mathcal N$. In this way, we could replace the pure state used to derive our equations with an {\it ensemble} of quantum systems that, instead of satisfying the deterministic evolution equations derived from the Hamiltonian constraint for the pure state \eqref{co-st}, satisfy a stochastic differential equation for their associated pure states with a density operator $\rho(\mathcal N,j)$  obeying a deterministic master equation of the Lindblad type \cite{Lindblad:1975ef}. By having such a description, the evolved distribution would then depend on the initial parameters $j_i$ and $\mathcal N_i$. Then, statistical analysis involving, for example, a fluctuation-dissipation theorem for quantum systems could allow for an exploration of the asymptotic regime and uncover the effective rescaling for $\bar{j}$. We leave the details of this proposal for future work.

\section{Discussion and concluding remarks}\la{sec:disc}

The notion that our observable universe could have emerged from within the interior of a black hole has its origin in the early seventies \cite{ PATHRIA}. A more concrete proposal for this scenario appeared a decade later in \cite{Frolov:1988vj, Frolov:1989pf},
where the hypothesis about the existence of a limiting curvature was used to glue the interior Schwarzschild region to a portion of de Sitter  by matching the two geometries at some space-like surface where the curvature reaches this  Planckian upper bound. 
 A specific example of this ``limiting curvature construction''  in 1+1 dimension was presented in \cite{Trodden:1993dm}. 
 Further motivation for pursuing this concept was provided in \cite{Easson:2001qf}, mainly 
 in relation to the classical problems of
 big bang cosmology (see also \cite{Poplawski:2010kb} for an implementation of this black hole cosmology scenario in the presence of spacetime torsion).

The Penrose diagram proposed in  \cite{Frolov:1989pf} constitutes an example of  a wider category of effective metrics delineating regular
solutions of the Einstein equations endowed with  an event horizon, or for short
``regular black holes'' (see \cite{Ansoldi:2008jw} and references therein). Included among these are the metrics describing black hole--white hole transition which have lately gained traction \cite{Hajicek:2001yd, Barcelo:2010vc, Haggard:2014rza, BenAchour:2020gon} and are derived in  polymer models  \cite{Chiou:2008nm, Ashtekar:2018cay, Kelly:2020uwj}. A distinguished characteristic for this class of geometries is the existence of an inner Cauchy horizon.  While for the Schwarzschild--de Sitter model of \cite{Frolov:1988vj} a stability argument was provided in
 \cite{Balbinot:1990zz}, the black hole--white hole model
 is affected by the well-known instability problem  of the inner horizon known as the ``mass inflation''  \cite{Poisson:1989zz, Brown:2011tv}  (see  \cite{Hamilton:2008zz} for a review on different aspects of this mechanism) as well as  the  issue of an infinite evaporation time that was more recently pointed out in  \cite{Carballo-Rubio:2018pmi}.
 
The  analysis  presented in this paper leads to the Penrose diagram shown in Fig. \ref{penrose} which differs from those that appear in the above proposals. In particular, no inner horizon is present in the effective spacetime region replacing the classical singularity and hence no white hole instability problem arises. Moreover, in the construction of  \cite{Frolov:1988vj} both de Sitter horizons are included and the Schwarzschild collapse is followed by a deflationary phase of the de Sitter spacetime before transitioning to an inflationary phase.
 On the other hand,  the numerical solutions of the effective dynamical equations derived in Section \ref{sec:eff-eq} show that the inflationary phase starts immediately after the bounce and it is asymptotically described by the top patch of the de Sitter spacetime shown in Fig. \ref{dS-static}, with no cosmological horizon appearing. In fact, the metric derived in Section \ref{sec:dS} belongs to yet another kind of a regular spherically symmetric black
holes that contains an expanding Kantowski--Sachs universe inside its event horizon and a spacelike scri in place of its singularity. These particular spacetimes had previously emerged in the literature and were dubbed
 ``regular phantom black holes'' in \cite{Bronnikov:2005gm} and  ``black universes'' in  \cite{Bronnikov:2006fu}. They were obtained as solutions to the Einstein equations sourced by spherically symmetric distributions of a scalar field called phantom matter. However, though the resulting Penrose diagrams are identical (see Figure 1.4b in  \cite{Bronnikov:2006fu}), our analysis does not require any exotic form of matter modeling the dark energy component of the universe. Instead, we provided a concrete realization of black hole singularity resolution with an emerging de Sitter universe by considering the quantum gravitational effects encoded in a set of effective equations derived from a quantum gauge fixed version of the full LQG framework. Such an explicit derivation from a given quantum gravity approach had been missing till now. The fact that dark energy is an emergent property of our effective spacetime and not an input, together with the long sought after consistency check for the physical relevance of the numerical value \eqref{Imm} of the Barbero--Immirzi parameter are the most striking consequences of our program thus far.


To gain a more profound physical intuition for the origin of the cosmological constant that appears in our effective geometry, it would be beneficial to contemplate on possible connections with the proposal of \cite{Josset:2016vrq, Perez:2017krv}. The authors in those papers proposed a mechanism  for the emergence of an effective dark energy from potential spacetime discreteness in quantum gravity. More precisely, noting that within the framework of unimodular gravity an energy-violation current can mimic a cosmological constant term in the Einstein's equations, a source of energy dissipation was modeled by friction-like forces acting on massive fields propagating on a granular spacetime. This effective description was aimed at capturing the interaction of matter degrees of freedom with the quantum discrete structure of geometry at the Planck scale. It resulted in a correct order of magnitude estimate for the cosmological constant that we currently observe. As our analysis so far does not account for any coupling with ordinary matter \footnote{The components of the effective stress-energy tensor defined in \eqref{Teff} can, however, be understood as describing the couplings to a form of emergent dark energy that is purely of quantum gravitational origin.}, it is not clear how one can draw a comparison between this effective description and our quantum corrected dynamics. At the same time, it is suggestive that in both cases the discrete nature of quantum gravity degrees of freedom play a crucial role in the emergence of a cosmological constant term. In our case, the apparent culprit is the Struve function in the effective Hamiltonian \eqref{heff-IV+CS} that is related to the graph structure on the 2-spheres foliating the spatial leaves in the interior region. This function acts as a dissipative dynamical term that breaks the time reversal symmetry of the interior effective geometry that would otherwise be present in the LQC models. It would be very interesting to investigate possible connections between the previously mentioned concept of ``friction'' and dark energy by studying the gravitational collapse of a massive shell in the QRLG framework. We leave this exploration to future works.

We conclude by pointing out that our derivation lends credence to the concept of ``cosmological natural selection'' proposed in \cite{ Smolin:1990us} which aims to explain the value of the dimensionless parameters that appear in particle physics and cosmology.



\section*{Acknowledgments}
E. Alesci and S. Bahrami are grateful for not only the previous support that they received from the NSF grant No. PHY-1505411, the Eberly research funds of Penn State, and the Urania Stott fund of Pittsburgh Foundation, but also for the privilege of receiving invaluable academic mentorship and guidance from Abhay Ashtekar.
D. Pranzetti acknowledges that this project has received funding from the European Union's Horizon 2020 research and innovation programme under the Marie Sklodowska-Curie grant agreement No 841923.
Research at Perimeter Institute is supported in part by the Government of Canada through the Department of Innovation, Science and Economic Development Canada and by the Province of Ontario through the Ministry of Colleges and Universities.

\begin{appendix}

\section{de Sitter metric in $\{\tau,x\}$ coordinates}\la{App:deSitter}

\begin{figure}[h!]
\centering
\includegraphics[width=0.4\textwidth]{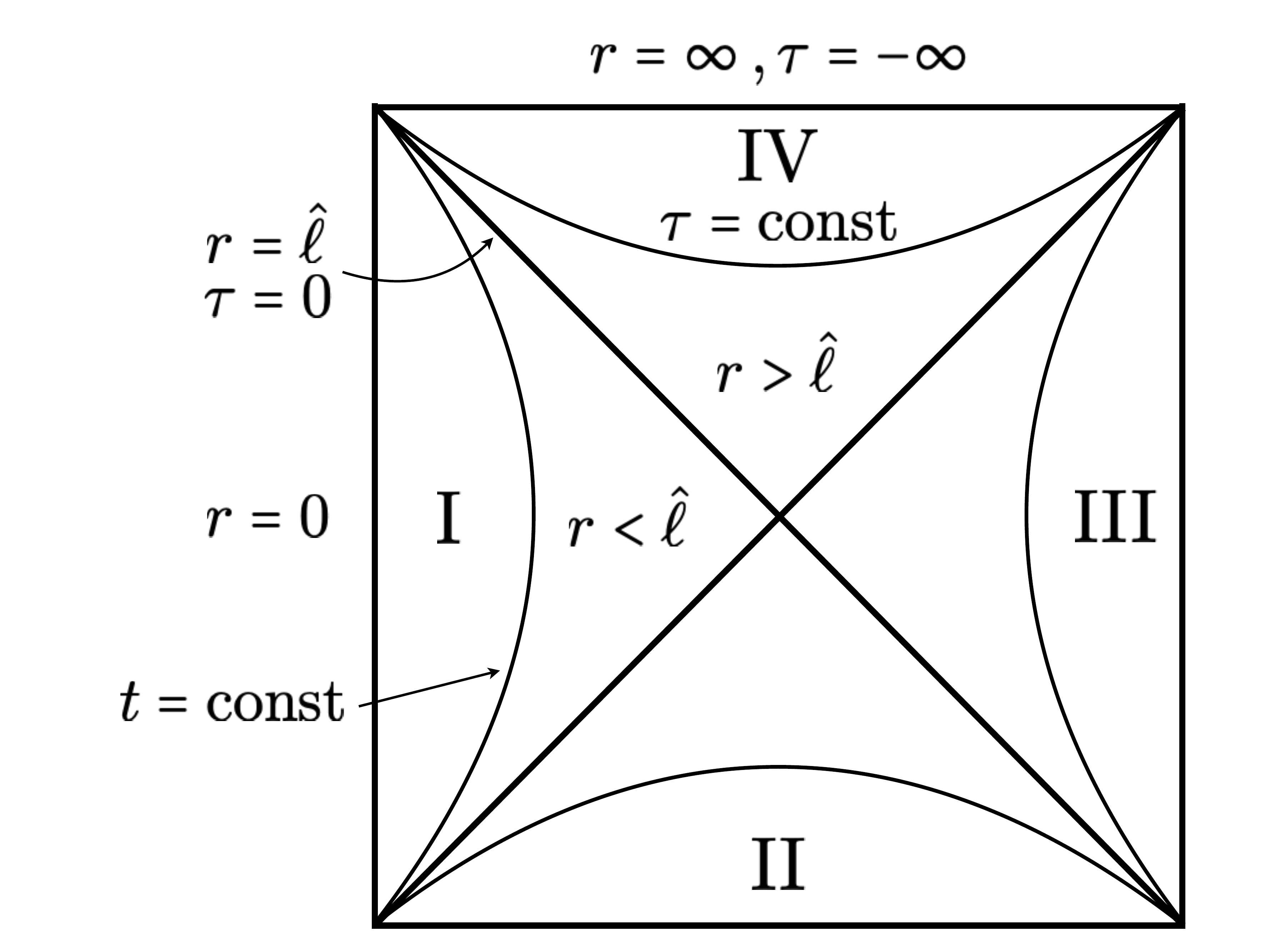}
\caption{de Sitter spacetime in static coordinates. The coordinates $r,t$ cover the  static regions I and III, where $r<\hat \ell$ and they are singular at the  cosmological horizons where $r=\hat \ell$. In regions II and IV, the spacetime is no longer static and we can introduce the proper time $\tau$ defined in \eqref{tau}. In these two regions the metric takes the form \eqref{dS3}. }
\label{dS-static}
\end{figure}

The reader may benefit from a brief review of the de Sitter metric in two different coordinate systems. First, recall that the de Sitter metric in the co-called ``static coordinates'' $r,t$ can be written as
\be\la{dS}
g_{ab}dx^a dx^b =-\Big(1- \frac{r^2}{\hat \ell^2}\Big) dt^2 +\frac{dr^2}{1-\frac{r^2}{\hat \ell^2}}   + r^2 d \Omega^2.
\ee
Here $\hat \ell$ is the cosmological length-scale associated with the cosmological constant that is given by $\lambda = 3/\hat \ell^2$. 
In regions I and III where $r<\hat\ell$ and $\partial_t$ is time-like, the metric \eqref{dS} is static. The cosmological horizons are located at $r=\hat\ell$. The static coordinates can be continued beyond these two horizons into  regions II and IV where $r>\hat\ell$ (see Fig. \ref{dS-static}). Here $\partial_t$ becomes  space-like and it is convenient to introduce 
 a new coordinate system:
\be \la{tau}
x=t\,,\quad \int N(\tau) \ d\tau = \int \frac{d r}{\sqrt{\frac{r^2}{\hat \ell^2}-1}} = \hat \ell \log{\Big[\frac{r}{\hat \ell}+ \sqrt{\frac{r^2}{\hat \ell^2}-1}\Big]}\,.
\ee
Here $N$ is an arbitrary function of $\tau$. 
The equation in the right-hand side can be inverted
to give  
\be 
r = \frac{\hat \ell}{2}\Big(e^{\int N(\tau) d\tau/\hat \ell}+ e^{- \int N(\tau) d\tau/\hat \ell}\Big) = \hat \ell \cosh{\left(\frac{\int N(\tau) d\tau}{\hat \ell}\right)}.
\ee
The metric \eqref{dS} in $\{\tau,x,\theta,\phi\}$ coordinates becomes  
\be \la{dS2}
g_{ab}dx^a dx^b = - N(\tau)^2 d \tau^2 + \sinh^2{\left(\frac{\int N(\tau) d\tau}{\hat \ell}\right)} \ dx^2 + \hat \ell^2 \cosh^2{\left(\frac{\int N(\tau) d\tau}{\hat \ell}\right)} \ d \Omega^2\,.
\ee
In the particular case where $N$ is a constant, we introduce the rescaled length-scale $ \ell= \hat\ell/N$ and the metric becomes
\be \la{dS3}
g_{ab}dx^a dx^b = - N^2 d \tau^2 + \sinh^2{\left(\frac{ \tau}{\ell}\right)} \ dx^2 +N^2 \ell^2 \cosh^2{\left(\frac{\tau}{\ell}\right)} \ d \Omega^2\,,
\ee
with the cosmological constant $\lambda$ given by
\be
\lambda=\frac{3}{N^2\ell^2 }\,.
\ee  

For our analysis, it is region IV of the Penrose diagram in Figure \ref{dS-static} that is relevant, as the effective metric we derive in Section \ref{sec:dS} approaches the metric \eqref{dS3} in the limit $\tau\rightarrow-\infty$.

\section{Deriving the asymptotic series solution}\la{sec:AppB}

We showed in Section \ref{sec:series} that an asymptotically Schwarzschild--de Sitter metric can be found by tuning most of the quantum parameters. Here we show that the converse is also true: assuming that $\xi$, $\gamma$, $\delta$, $\delta_x$, and $\nu$ are {\it given} as the solutions to Eqs. \eqref{PLorder}, \eqref{xi1}, \eqref{xi2}, \eqref{Hz2}, and \eqref{Lz2}, then the resultant metric is asymptotically Schwarzschild-de Sitter. Note that we are {\it not} assuming that the previously mentioned equations are derived from a certain order approximation of the dynamical equations. 

Our working assumption is that the phase space variables assume the following asymptotic forms:
\ba \label{asymp-series-gen} 
&& \lim_{z \rightarrow \infty} \Lambda(z) = \lambda_0 z + \lambda_1 + \frac{\lambda_2}{z} + \mathcal{O}(z^{-2}), \n\\
&& \lim_{z \rightarrow \infty} R(z) = \rho_0 z + \rho_1 + \frac{\rho_2}{z} + \mathcal{O}(z^{-2}), \n\\
&&\lim_{z \rightarrow \infty} P_\Lambda(z)= L_0 z^2 + L_1 z + L_2 + \mathcal{O}(z^{-1}), \n\\
&&\lim_{z \rightarrow \infty} P_R (z) = R_0 z^2 + R_1 z + R_2 + \mathcal{O}(z^{-1}).
\ea
While more complicated solutions are certainly possible, the leading pieces of the above expansions are borne out by detailed numerical investigations \footnote{We arrived at the same leading order behavior for $R$, $\Lambda$, $P_R$, and $P_\Lambda$ after imposing Eqs. \eqref{xi1} and \eqref{xi2} in \cite{Alesci:2019pbs}. Our new Hamiltonian converges to the one used in \cite{Alesci:2019pbs} as $R \rightarrow \infty$.}. Thus, we only consider the case where all four leading constants are non-zero.

Before delving into the equations, let us exploit a remaining gauge freedom to cast \eqref{asymp-series-gen} into a simpler form by fixing the coefficient of $z^{-1}$ in $R$. Sending $z \mapsto \sqrt{\rho_2/\rho_0} z$, 
we rewrite \eqref{asymp-series-gen} as 
\ba \label{asymp-series-1} 
&& \lim_{z \rightarrow \infty} \Lambda(z) =  \tilde{\lambda}_0 z \Big[1 + \frac{\tilde{\lambda}_1}{z} + \frac{\tilde{\lambda}_2}{z^2} + \mathcal{O}(z^{-2}) \Big], \n\\
&& \lim_{z \rightarrow \infty} R(z) =  \tilde{\rho}_0 z \Big[1 + \frac{\tilde{\rho}_1}{z} + \frac{1}{z^2} + \mathcal{O}(z^{-2}) \Big], \n\\
&&\lim_{z \rightarrow \infty} P_\Lambda(z)= \tilde{L}_0 z^2 + \tilde{L}_1 z + \tilde{L}_2 + \mathcal{O}(z^{-1}), \n\\
&&\lim_{z \rightarrow \infty} P_R (z) = \tilde{R}_0 z^2 + \tilde{R}_1 z + \tilde{R}_2 + \mathcal{O}(z^{-1}).
\ea
We emphasize that we are not assuming any a priori relation among the above constants. We shall drop tilde from these to simplify notation. 

The order $z$ terms in Eqs. \eqref{Rdot-CS} and \eqref{PLdot-CS} become
\ba \label{appz-1}
&& \xi + \cos{\Big[\frac{G \beta \gamma (L_0 \lambda_0 - R_0 \rho_0)}{\lambda_0 \rho_0 ^2}\Big]}=0, \hspace{3cm} \frac{L_0}{\rho_0 ^2}- \frac{\pi -\arccos{[\xi]}}{G \beta \gamma}=0.
\ea
Solving for $\xi$ and $L_0/\rho_0 ^2$, we find that the order $z^0$ terms in the scalar constraint equation and Eq. \eqref{Ldot-CS} coincide with Eqs. \eqref{xi1} and \eqref{xi2} respectively, which are satisfied by our original assumption.

Moving on to the order $z^0$ term in Eq. \eqref{Rdot-CS}, we find
\be
\rho_1 \bigg[L_0 + \frac{\rho_0 ^2 \Big(- \pi + \frac{\xi}{\sqrt{1- \xi^2}}+ \arccos{[\xi]}\Big)}{G \beta \gamma}\bigg]=0. 
\ee
Since by assumption $\rho_0 \neq 0$, Eq. \eqref{appz-1} implies that the square bracket  cannot vanish unless $\xi=0$, which is forbidden by the requirement that $\ell < \infty$. Therefore, this equation is satisfied if and only if $\rho_1 = 0$. Requiring this, the order $z^0$ term in Eq. \eqref{PLdot-CS} reduces to
\be \label{PL1}
\frac{G \beta \gamma}{\lambda_0 \rho_0} \big(L_1 \lambda_0 - R_1 \rho_0 + R_0 \lambda_1 \rho_0 \big) \sin{\Big[\frac{G \beta \gamma (L_0 \lambda_0 - R_0 \rho_0)}{\lambda_0 \rho_0 ^2}\Big]}=0.
\ee
Let us for now ignore the possibility that the sine function can vanish. This way, we must require
\be \label{L1}
L_1 \lambda_0 - R_1 \rho_0 + R_0 \lambda_1 \rho_0 = 0.
\ee
Taking the above, Eq. \eqref{appz-1}, and the vanishing of $\rho_1$ into account, the $z^0$ term in the scalar constraint equation requires the following
\ba 
&& L_1 \bigg[\pm \frac{\sqrt{2 \pi}}{8 \gamma} \sqrt{1-\xi^2} \big(2 \cos{[\iota]} + \pi h_{-1} [\iota]\big) + \pi \big(\sin{[\iota]} h_{-1}[\iota] + \cos{[\iota]} h_0 [\iota]\big)\bigg]=0, 
\ea
where $\iota := G L_0 \alpha \gamma/\rho_0 ^2 = \sqrt{2 \pi} (\pi - \arccos{[\xi]})/(8 \gamma)$. Since $\iota$ and $\xi$ are fixed by Eqs. \eqref{xi1} and \eqref{xi2} and the above square bracket is not implied by these two equations, we must require 
\be 
L_1 = R_1  - R_0 \lambda_1  = 0.
\ee
Considering the above derivations, the order $z^0$ term in Eq. \eqref{Ldot-CS} simplifies to $\lambda_1 = 0$, which then implies that $R_1=0$. In sum, by requiring Eqs. \eqref{xi1} and \eqref{xi2} we were able to deduce with no difficulty that $\rho_1=\lambda_1=L_1=R_1=0$. 

Before proceeding to the subsequent order, let us discuss the case where the sine function in 
Eq. \eqref{PL1} vanishes. This implies that $\xi =  1$. It follows from Eq. \eqref{xi1} that 
\be 
\sin{[\iota]} h_0 [\iota] = 0.
\ee
Hence, either $\sin{[\iota]}$ or $h_0[\iota]$ must vanish, but not both as that is only possible if $\xi = -1$. If $\sin{[\iota]} = 0$, then Eq. \eqref{xi2} requires $-2-\cos{[\iota]} = 0$ which cannot be. On the other hand, if $h_0[\iota]=0$, then the same equation requires $-2-\pi h_{-1}[\iota]/2=0$ which also cannot be satisfied for any value of $\iota$. Therefore, we must have $\sin{\Big[\frac{G \beta \gamma (L_0 \lambda_0 - R_0 \rho_0)}{\lambda_0 \rho_0 ^2}\Big]} \neq  0$. 

Moving on to the order $z^{-1}$ term in Eq. \eqref{PLdot-CS} and setting $\rho_1=\lambda_1=L_1=R_1=0$, we find
\ba \label{PLz-1}
&& 2 L_0 - L_2 + \frac{2 \xi}{G \beta^3 \gamma \delta \delta_x \sqrt{1- \xi^2} \big(2 \sin{[\iota]}+ \pi h_0 [\iota]\big)} \bigg[ \pi h_0 [\iota] \Big(-4 \ell_p ^2 \beta^2 \gamma^2 \delta + \delta_x \Big[ \ell_p ^2 \gamma^2 \big(-4 (-3 + \nu) \beta^2 + \ell_p ^2 \delta \big)+\beta^2 \delta \rho_0 ^2 \Big] \Big)\n\\
&& + \sin{[\iota]} \Big(-8 \ell_p ^2 \beta^2 \gamma^2 \delta + \delta_x \Big[\ell_p ^2 \gamma^2 \big(8 (-1+3 \nu) \beta^2 + \ell_p ^2 \delta \big)+ 2 \beta^2 \delta \rho_0 ^2\Big]\Big)=0.
\ea
Solving the above for $2 L_0 - L_2$ and assuming that $\sin{\Big[\frac{G \beta \gamma (-L_0 \lambda_0 + R_0 \rho_0)}{\lambda_0 \rho_0 ^2}\Big]}>0$ \footnote{If $\sin{\Big[\frac{G \beta \gamma (-L_0 \lambda_0 + R_0 \rho_0)}{\lambda_0 \rho_0 ^2}\Big]}<0$, then the phase space variables no longer admit the asymptotic form given in Eq. \eqref{asymp-series-1}. See the discussion at the end of Sec. \ref{sec:series}.}, the order $z^{-1}$ term in Eq. \eqref{Rdot-CS} reduces to the following simple equation:
\be \label{b11}
R_0 (1+ \lambda_2) - R_2 = 0.
\ee
Now let us use this last equation to simplify the order $z^{-2}$ term in the scalar constraint equation. If we impose Eq. \eqref{Hz2}, we end up with 
\ba \la{B12}
&&(2 L_0 - L_2)\bigg[- \frac{\sqrt{2 \pi (1- \xi^2)}}{8 \gamma} \big(2 \cos{[\iota]} + \pi h_{-1}[\iota]\big)- \xi \big(2 \sin[\iota] + \pi h_0 [\iota]\big) - \pi \big(\sin{[\iota]} h_{-1}[\iota]+ \cos{[\iota]} h_0 [\iota]\big) \bigg]=0.
\ea 
As the above square bracket is not implied by Eqs. \eqref{xi1} and \eqref{xi2}, we must require $2 L_0 = L_2$. Inserting this in Eq. \eqref{PLz-1} and subtracting off the last equation in \eqref{PLorder} we find that $\rho_0 = \pm G m N_0/\xi$.

Finally, taking the above simplifications into account, the order $z^{-2}$ term in Eq. \eqref{Ldot-CS} becomes 
\ba 
&&32 \beta^2 \delta \delta_x \lambda_2 G^2 m^2 N_0 ^2 \big(2 \sin{[\iota]}+ \pi h_0 [\iota]\big)^2 + 4 \pi \sqrt{2 \pi (1- \xi^2)} \ell_p ^2 \gamma \xi \delta_x [32 \beta^2 (-1+\nu) - \ell_p ^2 \delta] \big(\sin{[\iota]} h_{-1}[\iota] \n\\
&&- \cos{[\iota]} h_0 [\iota] \big)- \xi h_{-1}[\iota] \Big(768 \pi \ell_p ^2 \beta^2 \gamma^2 \delta \sin^2{[\iota]}- \delta_x \Big[128 \pi \ell_p ^2 \beta^2 \gamma^2 (1+\nu)+ \pi^2 \beta^4 \delta - 1024 \ell_p ^4 \gamma^4 \delta - 128 \ell_p ^2 \gamma^2 \n\\
&& \times [\pi (1+ \nu) \beta^2 - 8 \ell_p ^2 \gamma^2 \delta] \cos{[\iota]}\Big]\Big)- 32 \ell_p ^2 \gamma^2 \xi^2 \big(2 \sin{[\iota]} + \pi h_0 [\iota]\big) \Big(\pi h_0 [\iota] \big(-4 \beta^2 \delta + \delta_x [-4 (-3 + \nu) \beta^2 + \ell_p ^2 \delta]\big)\n\\
&& + \sin{[\iota]} \big(-8 \beta^2 \delta + \delta_x [8 \beta^2 (-1+3 \nu) + \ell_p ^2 \delta]\big) \Big)+ 2 \pi \xi \cos{[\iota]} \Big(\beta^4 \delta \delta_x - 64 \ell_p ^2 \gamma^2 h_0 ^2 [\iota] \big[3 \pi \beta^2 \delta - \delta_x \big(\pi (1+\nu) \beta^2 \n\\
&& - 8 \ell_p ^2 \gamma^2 \delta \big)\big]\Big)=0.
\ea
Using the asymptotic value for $N_0$ and imposing Eqs. \eqref{PLorder} (the last one) and \eqref{Lz2}, we find $\lambda_2 = -1$. Eq. \eqref{b11} then gives $R_2=0$. 

As expected, we have ended up with a 2-parameter family of solutions labeled by $\lambda_0$ and $R_0$. We started with four constants of integration, one for each first order ODE. That number was reduced to three due to the Hamiltonian constraint and then to two due to the remnant gauge freedom in rescaling the time coordinate $\tau$. 

Let us summarize what we have demonstrated here: if we tune $\xi$, $\gamma$, $\delta$, $\delta_x$ and $\nu$ using Eqs. \eqref{PLorder}, \eqref{xi1}, \eqref{xi2}, \eqref{Hz2}, and \eqref{Lz2}, then as long as the phase space variables are described asymptotically by the Laurent series expansions in $z^{-1}$ given in Eq. \eqref{asymp-series-1}, the interior metric is necessarily asymptotically Schwarzschild-de Sitter. The significance of this result is in the fact that the details of the initial data that we impose on $\Sigma$ are largely irrelevant so long as  Eq. \eqref{asymp-series-1} is valid. Our numerical investigations in \cite{Alesci:2019pbs} suggest that this is indeed the case for small variations around the Schwarzschild data \footnote{There we had focused on black holes with $m \gg m_p$. In that case, the classical phase space functions solve $\mathcal{H}^{\rm IV+CS} _{\rm cor} = 0$ to an arbitrarily high accuracy as $\Sigma$ tends to the event horizon.}. 
While we do not explicitly show, this solution is most likely an attractor for the Hamiltonian system under study.

\end{appendix}

\bibliography{Reference.bib}

\end{document}